\documentclass[11pt,showpacs]{article}  
\usepackage[paper=letterpaper,margin=1.0in]{geometry}

\usepackage{amssymb,amstext,amsmath,amsthm}
\usepackage[dvips]{graphicx}
\usepackage{latexsym}
\usepackage{psfrag}
\usepackage{amsfonts}
\usepackage{bm} % bold math
\usepackage{color}

\newcommand{\be}{\begin{equation}}
\newcommand{\ee}{\end{equation}}
\newcommand{\barray}{\begin{array}}
\newcommand{\earray}{\end{array}}
\newcommand{\bea}{\begin{eqnarray}}
\newcommand{\eea}{\end{eqnarray}}
\newcommand{\bs}{\begin{subequations}}
\newcommand{\es}{\end{subequations}}
\newcommand{\balign}{\begin{align}}
\newcommand{\ealign}{\end{align}}
\newcommand{\equ}{\begin{equation}}
\newcommand{\nequ}{\end{equation}}
\newcommand{\eqa}{\begin{eqnarray}}
\newcommand{\neqa}{\end{eqnarray}}

\def\nn{\nonumber}
\def\lb{\label}
\def\ci{\cite}
\def\pa{\partial}

\newcommand{\Ref}[1]{(\ref{#1})}

\newcommand{\mrm}[1]{\quad \mathrm{#1}\quad}

\def\om{\omega}

\let\eps=\epsilon

\def\d{\delta}

\def\g{\gamma}
\def\e{\epsilon}

\def\dd{\! \cdot\!}
\def\brd{\bm\rd}

\newcommand{\rd}{\mathrm{d}}
\newcommand{\p}{\partial}
\newcommand{\N}{\nabla}
\def\cL{ {\cal L}}

\def\f{\frac}

\def\bs{\bar{s}}

\def\n{\bm{n}}
\def\t{\bm{t}}

\def\s{\bm{s}}

\def\v{\bm{v}}

\def\ba{\bm{a}}

\def\bom{\bm{\omega}}

\DeclareMathOperator{\tr}{Tr}

\def\ts{{\bm{t^{*}}}}

\begin{document}

\title{  Non-equilibrium thermodynamics of gravitational screens}
\author{{Laurent Freidel$^{a}$\footnote{lfreidel@perimeterinstitute.ca}
 and Yuki Yokokura$^{b}$\footnote{yuki.yokokura@yukawa.kyoto-u.ac.jp}}
\\
%\smallskip \\ 
{\small ${}^a$\emph{Perimeter Institute for Theoretical Physics, }}\\
{\small \emph{
31 Caroline St. N, ON N2L 2Y5, Waterloo,Canada}} \\
{\small ${}^b$\emph{Yukawa Institute for Theoretical Physics, Kyoto University, Kyoto 606-8502, Japan}}\\
}
\date{}

\maketitle
\begin{abstract}
We study the Einstein gravity equations projected on a timelike surface,
which represents the time evolution of what we call a \textit{gravitational screen}. 
We show that  
such a screen behaves like a viscous bubble with a surface tension and an internal energy, 
and that the Einstein equations take the same forms as non-equilibrium  thermodynamic equations for a viscous bubble. 
We provide a consistent dictionary between gravitational and thermodynamical variables.
In the non-viscous cases there are three thermodynamic equations which characterize a bubble dynamics: These are 
the first law, the Marangoni flow equation and the Young-Laplace equation. 
In all three equations the surface tension plays a central role: 
In the first law it appears as a work term per unit area, 
in the Marangoni flow its gradient drives a force, 
and in the Young-Laplace equation it contributes to a pressure proportional to the surface curvature. 
The gravity equations appear as a natural generalization of these bubble equations 
when the bubble itself is viscous and dynamical.  In particular, it
shows that 
the mechanism of entropy production for the viscous bubble is mapped onto the production of gravitational waves.
We also review the relationship between surface tension and temperature, 
and discuss black-hole thermodynamics. 
\end{abstract}
%%%%%%%%%%%%%%%%%%%%%%%%%%%%%%%%%%%%%%%%%%%%%%%%%%%%%%%%%%%%%%%%%%%%%%%%%%%%%%%%%%%%%%%%%%%%%%%%%%%%%%%%%%%%%%%%%%%%%%%%%%%%%%%%%
\section{Introduction}
In trying to construct a quantum theory of gravity one is often led to wonder ``what are the fundamental space-time degrees of freedom or what are the quantum gravity constituents if any?''
 Such  questions are related to 
one of the crucial open questions in semi-classical gravity, that is:
``What is the entropy of a gravitational system?''.
All the recent data \cite{Planck} point towards having an early universe which is in perfect 
thermodynamical equilibrium as far as the matter fields are concerned.
In order to reconcile this fact with the second law, one needs,
as emphasized by Penrose \cite{Penrose}, to assign an entropy to the gravitational field
which roughly measures the amount of gravitational waves present in the space-time.
The open and pressing  problem is to understand ``what is the exact form of this gravitational entropy measure?''.
In order to address such a problem, 
we discuss the gravitational properties of a spacetime region.
We focus here on the situation where the boundary of this region is 
timelike and we call this boundary the  \textit{gravitational screen}.
And we investigate the gravitational dynamics from the point of view of this boundary. 
The reason we can, in the context of gravitational system, restrict our attention to the boundary of the region is 
ultimately due to the fact that the bulk energy is a constraint, and 
hence the canonical energy of a gravitational system is given by a boundary term \ci{W,B-Y, L}.

A typical example of a null gravitational screen is the black-hole horizon. 
It is well accepted that, in the case where the space-time is stationary and contains black-holes, the
Bekenstein-Hawking entropy \cite{BekH,H}  provides a measure of gravitational entropy: 
i-e the sum of the horizon's area in Planck unit is a measure of gravitational entropy \cite{G-H}. 
 
In the case where dynamics is present and the gravitational system is not at equilibrium, however, 
there are no accepted answers for the measure of gravitational entropy.
One of the challenges lies into the concept of horizons itself, that is 
``which timelike or null surface do we consider in the dynamical context to be the carrier of gravitational entropy?": 
the event horizon? which depends on the end of time, is therefore teleological and cannot be characterized by quasi-local measurements; 
the trapping or dynamical horizon? \cite{dynamicalH, dynamicalH2}  which is a local concept but which leads to hypersurfaces that are generically spacelike hence acausal; 
or the stretched horizon? \cite{Price,Membrane} which is somehow free of the both issues, since it is locally defined and timelike, but defined only as an approximation, not as a precise mathematical object, and known to possess instabilities. 
In thermodynamics the concept of entropy can be defined for systems which are out of equilibrium, 
 as long as they are under local equilibrium. 
 The central notion in this context is the  entropy production which controls the relaxation towards equilibrium.
%For out of equilibrium systems 
For gravitational systems which are out of equilibrium, 
understanding what entropy is, therefore means that we need to understand the mechanism of entropy production 
(see \cite{Eling,Chirco,Y,Clifton} for earlier discussion). 

The ultimate goal of our research is twofold. 
First we want to establish
a precise dictionary between the gravitational dynamics projected on a timelike screen and non equilibrium thermodynamical concepts and equations. 
This is what we achieve here, and in particular, 
this allows us to identify a gravitational concept analog to the viscous entropy production term.

Then, we want to ultimately select the types of gravitational screens that  have  time evolution compatible with the second law. 
Only such timelike screens can be promoted to be physical objects generalizing the notion of black holes to any region. 
As we will see, this amounts from the thermodynamical side to find the equation of state and constitutive relations of the gravitational screen.
The analysis of this issue is postponed to a future publication \cite{LY2}.
The correspondence we provide here is an powerful and suggestive analogy and not yet a full physical correspondence.

Our ultimate motivation is, however, to develop such a correspondence  and deduce from it  identification  what the entropy production is in the gravity system by pulling back the thermodynamical analog.
Another key motivation for our work is to devise a study of such a correspondence for {\it finite} regions of space-time, regions that do not refer to the end of space by studying asymptotically flat or AdS regions,
or the end of time, like event horizons.

Our work is profoundly motivated by the black-hole membrane paradigm 
\cite{Price, Membrane, Damour} and can be viewed as 
a full-fledge realization of this idea. The difference is that we start from the onset with
screens whose evolution is timelike, instead of treating the equation as an approximation of 
the near horizon geometry of a black hole.
As such we can formulate the gravity-thermodynamics correspondence for any region, not necessarily near an event horizon, 
and we do not treat it as an approximation.
Our work is also related to  Verlinde's idea \cite{Verlinde}, 
in which the gravitational degrees of freedom inside the screen can be 
replaced by the boundary degrees of freedom on the screen. It can be viewed as a fully relativistic and proper formulation of these.

In a previous work \cite{L} we argued, in the context of an Hamiltonian analysis,
that a gravitational screen  possesses a surface tension and an internal energy.
It was shown that the surface tension  is, in units where $8\pi G=1$, given by the inward radial screen's acceleration 
and the internal energy is given by minus  the screen radial expansion.
In this work we confirm %this analysis and show that 
this interpretation by showing that 
the Einstein equations projected on the time evolution of the gravitational screen  %reduce to thermodynamic equations for a viscous bubble.
take the same forms as the non-equilibrium thermodynamic equations for a viscous bubble. 
In the non-viscous cases the equations for a thermodynamic interface are of three kinds: 
The {\it first law}, the {\it Marangoni flow} equation and the {\it Young-Laplace} equation. 
In all three equations the surface tension plays a central although very different role: 
In the first law it appears as a work term per unit area, 
in the Marangoni flow its gradient drives a force, 
and in the Young-Laplace equation it contributes to a pressure proportional to the surface curvature. 
%In this work we show that 
In our analysis the gravity equations are turned out to correspond to a natural viscous 
generalization of these bubble equations, coupled to a  Newtonian gravitational potential.
This allows us  to give a %full 
consistent dictionary between thermodynamic entities and gravitational terms. 
Among other things it provides us with a characterization of the viscous 
entropy production  as a measure for the amount of gravitational waves flowing through  the screen.
One of the key point of our discussion,
which allow us to give a full dictionary, is the role of the surface tension and the Young-Laplace equation, which was missed in previous discussions and is essential for a quasi-local description.

Our work bares some resemblance with the  celebrated fluid-gravity correspondence developed by Minwalla et al \cite{Minwalla,Min2,Min3} in the context of asymptotic AdS spaces. In particular these authors recognized that some gravity equations are 
encoded in a relativistic conservation of energy and momenta, and reconstructed the space-time metric in a derivative expansion from a fluid solution.
The key difference is that the fluid-gravity correspondence refers to a screen at spatial
infinity in AdS space. The correspondence is between the relativistic fluid equation on this infinity screen
and the bulk metric in a long wavelength expansion.
Our correspondence, on the other hand, aims at a local reconstruction valid for arbitrary screens inside the manifold.
One might expect our quasi-local  correspondence to eventually agree with the fluid-gravity one in the limit of an asymptotic AdS screen.
Note however, that in this asymptotic limit there is no longer any reference to a Young-Laplace equation, which arises only in the finite boundary setting. The appearance of this equation within gravity is a new feature of our analysis.
It is also important to note that in the fluid-gravity correspondence the equations are autonomous, containing an expression for the constitutive law.
Even if our goal is to get a similar expression for finite screens, we do not achieve this in the current paper.

Another instance where gravity and thermodynamics are put together in a local manner is 
in the work of Jacobson \cite{Jacobson}, which aimed at identifying  Einstein equations with an equation of state. 
There, it is postulated that the radial acceleration {\it is} the temperature, based on Unruh's argument \ci{Unruh}. 
%In Jacobson's approach, however, it is postulated that the radial acceleration {\it is} the temperature.
However, this discussion %derivation of the Einstein equaiton from a thermodynamic equation 
is valid only %in static space-times or equilibrium configurations but is it when we have only local equilibrium?
for a special screen where the expansion of null rays vanishes and where the flux of gravitational waves going through the screen can be neglected. 
In contrast, what our analysis clearly shows is that, for a general screen in a  \textit{general} spacetime which includes  dynamical situations, 
the radial acceleration should be interpreted as the surface tension, not a temperature. 
Note that our argument comes from the appearance of the radial acceleration  in three distinct gravity equations which can be interpreted as thermodynamic ones. 
Our discussion, on the other hand, does not include a quantum discussion which is necessary in order to  identify the concept of local equilibrium, and especially, the  temperature associated to a general screen. 
Therefore, our analysis is still missing one central ingredient, 
and in this sense our dictionary is, for now at least, analogical. 
However, our strategy, that is, one which starts from the identification of non-equilibrium thermodynamic equations, 
is geared towards the discovery of the entropy of spacetime. 
That is because %in such a case where we still don't know what form the entropy takes,
the focus  on the second law, that is, the mechanism of entropy production in the context of the thermodynamical analogy we are developing,  
should be effective for finding the definition of the entropy density
\footnote{This approach is different from the usual idea of fluid dynamics, 
in which one starts from the local equilibrium condition, where the entropy density is introduced, 
considers fluctuations from it, and determines dynamics according to the second law \ci{Groot}. 
However, 
based on modern technique of non-equilibrium statistical mechanics, 
the hydrodynamic equations can be derived by starting with construction from microscopic dynamics
of a quantity which increases in dynamical processes \ci{Sasa}.

Note also that our attempt does not depend on conditions such as black-hole horizon and null-limit as discussed in \ci{Vanzo}, 
and it is also not restricted to the use of global concepts and some adiabatic condition as in \ci{Barcelo}.}. 
Furthermore, one should keep in mind that such a general concept of temperature is not yet fully developed 
%the concept of temperature for a general space-time screen is not yet fully developed 
(see however \cite{matteo} for an interesting proposal in the context of 2d space-time). 

At the end of this paper 
we present an understanding that explains the tension between 
interpreting the radial acceleration as a surface tension as we do 
versus interpreting it as a temperature as is often done. 
In thermodynamics of an interfacial system 
the relationship between surface tension and temperature is given by the Gibbs relation \ci{Molecule,PCS}. 
Using this we try to understand the black-hole thermodynamics from our framework. 
Note however that this discussion is analogical and speculative, 
since we have not established the physical identification of the surface tension yet. %the black hole yet. 

%In this paper we also present a new understanding that explains the tension that exists between interpreting the radial acceleration as a surface tension as we do versus interpreting it as a temperature as it is often done.
%In thermodynamics of a interfacial system the relationship between surface tension and temperature is given by the Gibbs relation, 
%which we describe at the end of this paper.
%This relation stipulates that the derivative of the surface tension with respect to the temperature is given by the entropy per unit area.
%So the only instance where the surface tension depends linearly on the temperature is when the entropy per unit area of the interface is  constant and independent of the temperature. 
%This is realized in the case of the stationary black hole, and we can understand the black-hole thermodynamics from our framework. 
%Whether we can assume that the entropy per unit area of certain gravitational screens 
%is constant even in the non-equilibrium context still needs to be demonstrated.

There are other approaches which generalize the black-hole membrane paradigm. 
These differ from our perspective and none puts emphasis on the Young-Laplace equation as a key equation for the correspondence (with the exception of \cite{YL}).
In \ci{Bredberg:2011jq} the incompressible Navier-Stokes equation on a timelike surface at any position
is derived from a limit of the Einstein equation by using 
a flat uniformly accelerated screen and a long wavelength perturbation 
around a static metric with horizon. This is not valid  for a general spacetime unlike our treatment. 
\ci{Padmanabhan:2010rp} analyzes any null hypersurface in the freely-falling frame, 
and the equation becomes the Navier-Stokes equation of a usual Stokes fluid. 
In \ci{G1, G2, G3} a quasi-local analysis similar and complementary
 to one we present here, has been developed
to derive fluid-dynamics-like equations for any hypersurface.
This formulation uses null vectors unlike our formulation to define the thermodynamic
 quantities, and the corresponding dictionary differs in details.
This analysis  identifies correctly a notion of internal energy  as  the expansion
and entropy production as gravity wave production but it does not identify the surface tension 
as one of the key ingredients and the corresponding Young-Laplace equation as one of the  key thermodynamic equation.
 
Our work is also  inspired by the developments associated 
with trapping, isolated, dynamical and slowly evolving horizon \ci{dynamicalH,dynamicalH2,Hayward, Ashtekar2, BF}.  
These are not timelike  however, and therefore, they cannot be interpreted as a 
 membrane supporting a real physical system.
Finally, in the context of spacetime thermodynamics, 
non-equilibrium effects have been considered by analysing the null Raychaudhuri equation, 
which is  restricted to cases where the expansion vanishes \ci{Eling,Chirco,Y}. In this analysis the entropy production has been discussed and agrees in a null limit with our interpretation.

Since this paper is mainly devoted to a gravity  community, we review in the beginning  elements of fluids dynamics and give the main concepts and equations.
We then also present the thermodynamics of interfaces and describes some element of the physics of surface tension. This allows us to set the definitions and notations that we use in the following. The knowledgeable reader can jump directly to the next sections.
In these we present the $2+2$ decomposition of gravity,  introduce the membrane energy-momentum tensor and analyze the gravity equation projected on the screen's evolution.
We then give the consistent dictionary between gravity and thermodynamics. 
Finally we review the Gibbs relation between temperature and surface tension, 
and discuss the black-hole thermodynamics from this point of view.

%%%%%%%%%%%%%%%%%%%%%%%%%%%%%%%%%%%%%%%%%%%%%%%%%%%%%%%%%%%%%%%%%%%%%%%%%%%%%%%%%%%%%%%%%%%%%%%%%%%
\section{Non-relativistic Fluid dynamics}
In this section we recall the key elements of the laws of conservation 
for a viscous fluid consisting of only one component \ci{Groot,Landau_F}. 
We denote  the fluid velocity field at a given point $\bm x$ in time  by $v_{i}(t,\bm x)$. 
In the Lagrangian picture one follows  fluid particles, 
and the rate of change of physical quantities is due to the flow and the explicit time dependence.
 The  particular derivative  is given, for a local quantity $f(t,\bm x)$, by
\be
\rd_{t} f \equiv \partial_{t}f + \v \dd \brd f,
\ee
where $\brd$ is the spatial derivative in the laboratory frame which we are now considering.  Here and in the following  $\cdot$ denotes one index contraction and bold represents vectors or tensors.
Suppose that we have an integral over a domain $I =\int_{D} f $. Then
\be
\frac{\rd I}{\rd t} = \int_{D} (\partial_{t}f + \brd \dd ( f \v) ) 
=\int_{D} (\rd_t f + f \sigma)~,
\ee
where $\sigma \equiv \brd \dd \v$ is the compressibility of the fluid, which measures the expansion rate. 
We denote by $\rho$ the mass per unit volume of the fluid 
%the mass density of the fluid.
The mass conservation equation or continuity equation is 
\be
\partial_{t}\rho + \partial_{i} (\rho v^{i}) =0.
\ee
This can also be written in terms of the Lagrangian derivative as 
\be\label{Mass}
{\rd_{t}\rho + \sigma \rho  =0}~.
\ee
This equation will be used to rewrite all derivatives of %a mass density $\hat f$  
a quantity per unit mass $\hat f$
in terms of %its volume density 
its density per unit volume 
$f\equiv \rho \hat f$ as  $\rho \rd_{t} \hat{f}= \rd_{t}f +\sigma f$ \footnote{Note that we will call $f$ a density (per volume) while a hatted quantity is called a ``mass density'' (per unit mass).}.

A fluid within a domain $D$ is submitted to bulk and boundary forces. 
Here we suppose that the bulk forces are given by a conservative force 
$-\rho \brd \phi$, due to the presence of a Newtonian potential $\phi$ and by 
an external radiative force  $\bm f$ per unit volume.
In our context it is convenient to think of this external force as being due to radiation going through the system. This is even more adequate in the case the system is itself part of an interface as we will later assume.
Boundary forces are provided by the surface density force $ T_{ij}s^{j}$, 
where $s^{i}$ is the boundary unit normal and $T_{ij}$ the total stress tensor. 
Here $T_{ij}$ represents the flux in direction $j$ (rate of change per unit area) of momenta $i$.  
In summary  the total force  is: 
\be
F_{i} =\int_{D} {f}_{i} - \int_{D} \rho \partial_{i}\phi+\int_{\partial D} T_{ij}s^{j}.
\ee 
The conservation of angular momenta implies that,
in the absence of vortices, the total stress tensor $T_{ij}$ is symmetric.
If the fluid is non-elastic\footnote{For an elastic fluid the stress tensor at rest is not a scalar but an elastic tensor.},
the total stress tensor can be decomposed in terms of a pressure term $p$ and 
a viscous stress tensor $\Theta_{ij}$.
\be
T^{i}{}{}_{j}= -p \delta^{i}{}_{j} + \Theta^{i}{}_{j}.
\ee
The pressure $p$ represents the stress tensor when the fluid is at rest.
Note that in general, the dynamical pressure $P = p - \theta/3$, where $\theta\equiv \Theta^{i}{}_{i}$, 
differs from the static pressure $p$, unless the bulk stress $\theta$ vanishes, where the fluid is said to satisfy Stokes's assumption, 
which is valid for mono-atomic gases but not generally.

The fundamental equation that characterizes the conservation of momenta is given by 
\be\lb{NSeq}
\rho (\partial_{t}{v}_{i} + v^{j}\partial_{j}v_{i}) = \rho \rd_{t}v_{i} =f_{i} -\rho \partial_{i}\phi + \partial_{j}T^{j}{}_{i}~,
\ee
which is the Navier-Stokes equation for a general fluid since we have not imposed constituent equations yet. 
It will be convenient to rewrite this equation in term of the momentum density of the fluid 
\be
\pi_{i}\equiv \rho v_{i}.
\ee
The conservation of momenta reads 
\be\label{momenta}
{\rd_t \bm \pi +\sigma \bm{\pi} =-\brd p  + \brd \dd  \bm{\Theta} -\rho \bm{\rd}\phi + \bm{f}}.
\ee

Next we are interested in looking into the conservation of  energy.
The total energy density $\hat e$ per unit mass is given by the sum of kinetic energy $\f{1}{2}\v^2$, 
gravitational energy $\phi$, and internal energy density $\hat u$,
\begin{equation}
\hat e = \f{1}{2}\v^2 + \phi + \hat u~.
\end{equation}
The internal energy represents the energy of  thermal agitation 
plus the energy of  short range interactions. 
In the \textit{most general} case, the total energy changes in three ways: 
energy flux going through the boundary, time dependence of the Newtonian potential, 
and radiative heat transfer. 

Firstly, the energy flux comes from three contributions: 
It contains first a convective term $\rho \hat{e} \bm{v}$ due to the flow of matter, 
there is also a boundary work term by the surrounding fluid $\int_{\p D} (-T_{ij} v^i s^j)$, 
and finally there is the flow of heat incorporated in the heat flux vector $\bm{q}$, 
which captures how much heat enters into the system through the boundary $\p D$. 
Secondly, we have a 
contribution 
which comes from the time dependence of the potential energy $\rho \p_t \phi$.
Finally, we can lose energy due to the presence of a radiative energy transfer $\dot E_{\mathrm{rad}}$. 
Radiation with high frequency can enter into the bulk of the fluid to contribute directly to heat 
without converting into conductive heat transfer, such as a system under cosmic ray and a nuclear power reactor.
Thus we reach 
$
\p_t e =  \p_i(T^{ij}v_j - ev^i- q^i) + \rho \p_t \phi  + \dot E_{\mathrm{rad}}~,\nn
$
where $e=\rho \hat e$. This  can also be written as 
\begin{equation}\lb{energy}
{\rho \rd_t \hat e =  \p_i(T^{ij}v_j - q^i) + \rho \p_t \phi  + \dot E_{\mathrm{rad}}}.
\end{equation}

%%%%%%%%%%%%%%%%%%%%%%%%%%%%%%%%%%%%%%%%
%%%%%%%%%%%%%%%%%%%%%%%%%%%%%%%%%%%%%%%%%%%%%%%%%%%%%
\subsection*{The first law}
Lets construct the local version of the first law of thermodynamics.
First we can evaluate, from \Ref{NSeq} and the definition of $\rd_t$,
\bea\label{dvv}
\f{1}{2}\rho \rd_t \v^2&=& -\rho \v \dd \brd \phi + \bm{v}\dd \bm{f} + v_{i}\partial_{j}T^{ij}\nn \\
&=&-\bm{\pi} \dd \brd \phi+ \v\dd \bm{f} + \partial_{i}( T^{ij} v_{j}) - \bm{T} \!:\! \bm{\Sigma}
\eea
where we have introduced the  {\it strain rate }tensor $\Sigma_{ab} \equiv \partial_{(a}v_{b)}
%=\dot{D}_{ij}
$, %the parenthesis meaning symmetrization.
the parenthesis means symmetrization, and $\!:\!$ represents a double contraction. 
We also have that 
\begin{equation}\lb{dphi}
\rho \rd_t \phi = \rho \p_t \phi + \rho \v \cdot \brd \phi~.
\end{equation}
By subtracting these two equations from \Ref{energy}, 
we obtain 
\be
\rho \rd_t \hat{u} =  T_{ij} \Sigma^{ij}    - \partial_{i}q^{i} + \dot Q_{\mathrm{rad}}.
\ee
We  have introduced $\dot Q_{\mathrm{rad}}= \dot E_{\mathrm{rad}}- \v\dd \bm{f}$
which represents the amount of radiative heat  transferred to the system, i-e the total energy transfer minus the work terms due to radiation. 
Introducing the internal energy density $u\equiv \rho \hat{u} $, and using the decomposition $\bm{T}=-p \bm{\delta} + \bm \Theta$, 
we can write the local version of the first law

\begin{equation}\lb{1st}
{\rd_t u + \sigma u = -p \sigma + \bm \Theta : \bm \Sigma - \brd \cdot \bm q + \dot Q_{\mathrm{rad}}}.
\end{equation}
This result tells that effects of potential fields such as $\phi$ do not enter into the first law. 
We have introduced two forms of heat, the usual heat transfer across boundary and the radiative heat transfer. 
In the case that the system is itself part of an interface this distinction becomes even more crucial, 
since the system can transport heat not only across its boundary along the interface but also across the interface. 
Both forms of heat transfer can  contribute to the internal energy positively or negatively, 
depending on processes.

%%%%%%%%%%%%%%%%%%
%%%%%%%%%%%%%%%%%%
\subsection*{Gibbs relation and the second law}
If one assumes the condition of {\it local equilibrium}, that is equilibrium for the fluid particle even in a context where the fluid is out of equilibrium, we can define the notion of entropy density.
Let us recall that in the thermodynamic limit, it is assumed that the fluid infinitesimal element still contains a extremely large numbers of molecules so that it makes sense to talk about the local value of microscopic concepts such as entropy or internal energy.
For a one-component fluid the entropy density per unit mass $\hat{s}$ is considered 
to be a function of the internal energy density per unit mass $\hat u$ and the specific volume $v=1/{\rho}$, $\hat s = \hat s (\hat u, v)$. 
Its variation is given by the Gibbs relation: 
 $ T \delta \hat{s} = \delta \hat{u} + p \delta v$. 
 Using the mass conservation $\rd_t v = {\sigma} v$, we can write the Gibbs relation  as a time variation:  
\begin{equation}\lb{G_rel_t}
T \rd_t \hat s = \rd_t \hat u + p \sigma v~.
\end{equation}
If we denote by  $ s=\rho \hat{s}$ the entropy density per unit volume 
and use the first law (\ref{1st}) in the Gibbs relation, we obtain  the evolution equation for the entropy simply given by
\be
{T\left(\rd_t s  +\sigma s\right) = \bm{\Theta}: \bm{\Sigma}  - \bm{\rd}\dd \bm{q} + \dot{Q}_{\mathrm{rad}}}.
\ee
If one multiplies this equation by the inverse temperature $\beta\equiv 1/T$ 
we obtain, in the usual case where there are no radiative heat transfer,
the entropy evolution law evolution  in the form:
\be\label{ds}
 \rd_{t} s +\sigma s = - \brd\dd \bm{J}_{s} + \dot{s} 
\ee
where the entropy current $\bm{J}_{s} \equiv \beta{\bm{q}}$ is proportional to the heat flux and $\dot{s}$ is the local entropy production   given by
\be{
\dot{s} = \beta \bm{\Theta}: \bm{\Sigma} +{ \bm{q}\dd {\brd \beta}}
}\label{dots}
\ee
It expresses that for a one-component fluid there are two sources of entropy production.
One is due to heat conduction and the second one is due to the gradient of the velocity field leading to viscous flow.
It is important to note that each term is always the product of a flux term: $\bm{\Sigma}$ for the shearing flux 
or $\bm{q}$ for the heat flux, times a thermodynamic force: $ \bm{\Theta}$ for the viscous force or
${\brd \beta}$ for the temperature force.
As we are going to see the relations between forces and flux represents the constitutive relations.
The second law of thermodynamics demands that the entropy production  be positive, which is expressed by the inequality
\be
\dot{s} \geq 0.
\ee
This is the local form of the second law.
It is important to note that the second law is a restriction on the type of constitutive relations.

In order to get the familiar formulation of the second law, lets  integrate (\ref{ds}) over a finite region $V$ and denote the total entropy by $ S = \int_{V} s$,  while we denote the rate of exchange of heat per unit time by  $ \dot{Q} =-\int_{\partial V} \bm{q}\dd \s$ where $\s$ is the outward unit normal to $\partial V$.
We now look at a transformation taking place in a time interval and assume that the system is closed (it doesn't exchange matter with its environment) and that the temperature is constant in time and over space. 
We can now integrate (\ref{ds}) over $V$ and over the time interval and get 
\be
\delta S = \frac{\delta Q}{T} + \delta S_{i},
\ee
$\delta_{i} S\equiv \int_{t_{i}}^{t_{f}}\int_{V}\dot{s}
$ denotes the {\it internal} entropy produced inside the system.
The second law stipulates that it is always non-negative and therefore 
we can write it as 
\be
\delta S \geq \frac{\delta Q}{T}.
\ee
The term ${\delta Q}/{T} $ describes the entropy the closed system { exchanges } with its environment and  it can be positive or negative.

%%%%%%%%%%%%%%%%%%%%%%%%%%%%%%%%%%%%%%%%%%%%%%%%%%%%

\subsection*{Constitutive Laws}
By {\it definition} a fluid is a system for which the viscous stress tensor $\Theta_{ij}$ is a function of the strain (or deformation) rate tensor $\Sigma_{ij}=\partial_{(i}v_{j)}$ only. 
A relationship $\Theta_{ij}(\bm{\Sigma})$ is called a {\it constitutive law} of the fluid. It characterizes its thermodynamic nature.
A generalized fluid  possesses memory effects 
in which $\bm \Theta$ is not only a function of $\bm{\Sigma}$ but also  $\dot{\bm\Sigma}$, $\ddot{\bm\Sigma}$ etc.
A {\it Newtonian } fluid is a fluid without memory, for which the relationship is linear, that is, 
\bea\lb{Newton_fluid}
\theta= \xi \sigma,\qquad
\Theta_{ij} = 
2 \mu \Sigma_{ij} + \frac{(\xi-2\mu)}{3} \delta_{ij}\sigma~,
\eea
where $\xi $ is the bulk viscosity while  $\mu$ is the shear viscosity.
Other type of constitutive law can model systems like lava, mayonnaise or mud,
that behave like solid below a certain critical value of the stress and fluid above.
This  can be realized for instance by the Bingham plastic law, which involves a yield stress $\tau_{c}$ and a viscosity $\mu$:
\bea
\bm{\Sigma} &=&0  \mrm{if} \tr [(\bm{\Theta} -\theta \bm{1}) \bm\Theta ]< \tau_{c},\mrm{while}\nn\\
\bm\Theta &=& \left(2\mu + \frac{\tau_{c}}{\tr [(\bm{\Sigma} -\sigma \bm{1}) \bm \Sigma]} \right)
\bm \Sigma  \mrm{if} \tr [(\bm{\Theta} -\theta \bm{1}) \bm\Theta ]> \tau_{c}.
\eea
As we will see different non-Newtonian constitutive relations can be encountered in the gravity setting.
Another type of constitutive laws needed to described the system is a relationship between 
heat flux vector $\bm{q}$ and temperature $T$.
The simplest relation is Fourier's law: $\bm q = - \kappa \brd T~$, 
where $\kappa$ is a heat conduction coefficient. 

It is important to recognize that the constitutive laws have status which differs radically  from the fundamental conservation laws since they are phenomenological and fixed by experimentation.
With the same material but under different stress and external conditions the experimentation can set up different constitutive law. 
One of best examples is cooking, which is the art of changing the constitutive laws in a tasteful manner.
It is also significant to note that 
the second law of thermodynamics restrict the possible set of constitutive laws
 $\bm\Theta(\bm\Sigma)$ and $ \bm q(\beta)$ (see \cite{Landau_F,Romat,Bhattacha} for a general discussion). In the Onsagerian regime, where we assume a linear relationship, it demands the positivity of the viscosity coefficients $\mu,\xi$ and of the heat conduction coefficient $\kappa$.

Let us conclude by saying that
 for a one-component fluid,
 once we use the  constitutive relations, $\bm\Theta(\bm\Sigma)$ and $ \bm q(\beta)$, we have 6 unknowns: $(\rho,~p,~u,~\v)$.
We have 5 fundamental conservations equations, mass, 3-momenta and energy.
The last relation is provided by the state equation $s(u,\rho)$.

%%%%%%%%%%%%%%%%%%%%%%%%%%%%%%%%%%%%%%%%%%%%%%%%%%%%%%%%%%%%%%%%%%%%%%%%%%%%%%%%%%%%%%%%%%%%%%%%%%%%%%%%%%%%%%%%%%%%%%
\section{Thermodynamics of systems with interfaces}
The previous equations describe dynamics of the bulk of a fluid.
In the case where there are interfaces between two phases, 
the interface itself will behave as a thermodynamic system. 
The goal of this section is to present the interface's thermodynamic equations. 
Apart from the mass conservation, there are three equations. 
In addition to the {\it first law} expressing the balance law of internal energy on the surface 
and the so-called {\it Marangoni flow} equation expressing the conservation of tangential momenta on the surface, 
we also have an extra equation, that is, the {\it dynamical Young-Laplace} equation expressing the conservation of normal momentum across the surface 
and governing the expansion of the interface within its environment. 
The presence of this extra equation is tied up with the appearance of a new physical quantity entering the description of fluid interfaces: the {\it surface tension}.
which  plays a specific role in these three equations.

In this section we will make a short review of physics of interface \ci{dG} to familiarize ourselves with the basics concepts. 
In the next section it will be found that the gravitational case corresponds to a natural 
generalization of these equations, where the 
interfacial fluid is viscous and the interface is allowed to dynamically change its size.

%%%%%%%%%%%%%%%%%%%%%%%%%%%%%%%%%%%%
\subsection*{Surface tension}
As soon as  there are interfaces between two phases, such as two immiscible liquids, two kinds of materials, or two phases of the same fluid, 
a new physical quantity makes its appearance: The surface tension $\g$. 
Its physical origin is due to the Wan der Walls interactions between molecules. 
Molecules tend to attract each other and bind 
when the attractive interaction is stronger than thermal agitation. 
If the binding energy per molecule is $U<0$ inside the liquid, 
a molecule living at the surface has the energy $\sim U/2$ 
because it is surrounded only by  half its neighbours. 
Boundary molecules are therefore in excited states compared to the bulk ones.
Thus, any fluid  will adjust its shape in order to 
make the exposed surface area the smallest possible. 
The surface tension $\g$ is a direct measure of this energy cost per unit area.
If $a$ is the molecule's size and $a^2$ is its exposed area, the surface tension $\g$ is estimated as $\g\sim U/(2a^2)$.
Note that in general the surface tension $\g$ is determined by the relation between the two kinds of materials and phases across the interface, 
and that dimensionally $[\g]=(energy)\times (length)^{-2}$. 

The basic definition of surface tension is that \textit{$\g$ is the energy that must be provided to increase the surface area by one unit.}
In the equation this means that the amount of work $\delta W$ needed in order to 
increase the surface area by $\rd A$ is 
\begin{equation}\lb{W_g}
\d W = \g \rd A.
\end{equation}
This leads to an expression of the interfacial thermodynamic relation for internal energy.
If $U^{I}$ denotes the interfacial internal energy, $S^{I}$ the interfacial entropy
and $N^{I}$ the number of interfacial constituent,
the relation for a one-component fluid can be written as \ci{Molecule,PCS}
\be\label{interfU}
\rd U^{I} = T\rd S^{I} + \gamma \rd A + \mu_{I} \rd N^{I}
\ee
where $\mu_{I}$ is the interface chemical potential
\footnote{Here the idea of \textit{dividing surface} by Gibbs is used to treat the interface 
as a thin layer without volume in the limit \ci{Molecule,PCS}. 
Then, choosing the position of the dividing surface, we can put $N^I=0$ for a one-component fluid.}. 

We can see that this equation is similar to the thermodynamic relation of the bulk phases 
except that the 
work term involves an area variation. 
This suggests that we can interpret formally the surface tension as a negative surface pressure in a 2d system:
\be\lb{g=-p}
\gamma = -p_{2d}.
\ee
This interpretation follows from understanding the pressure as a repulsive force 
due to thermal agitation, and hence a binding force $\g$ can be accounted as a negative pressure.

%%%%%%%%%%%%%%%%%%%%%%%%%%%%%%%%%%%%%%%%%%
\subsection*{Marangoni flows}
The interpretation \Ref{g=-p} also suggests that 
gradient of surface tension will drive tangential flow in the opposite way a gradient of pressure would.
These flows are known as Marangoni flows.
The presence of a gradient of surface tension creates a Marangoni force $\bm  f=\brd \g$. 
This force induces a flow that {\it moves the interfacial fluid 
from region of low surface tension toward region of higher surface tension}.
If $\pi_A$ denotes the interfacial fluid momenta we have that
\be\label{marangoni}
\rd_t \pi_{A} = \rd_{A} \g~,
\ee
where the index $A,B,...$ represents components tangent to the interface, and 
$\rd_A$ is here the covariant derivative on the surface since the surface itself is the thermodynamic physical system we are considering.  

For usual materials a gradient of temperature $T$ results in a gradient of surface tension, 
and thus, a hotter fluid  has lower surface tension. Hence, $\partial \gamma /\partial T <0$, and  
the surface tension decreases until the critical point. 
Similarly, a gradient of concentration $c$ results in a gradient of surface tension. 
For a surfactant like soap, whose presence lowers the surface tension, 
we have that $\partial \gamma /\partial c <0$. 

A famous example of the Marangoni flow arises in the ``tears of wine'' effect, 
which leads to tears of wine continuously forming on the glass at a certain height above the wine surface on top of a film.
In a glass of wine  (mixture of water and alcohol), a film forms due to the fact that 
the surface tension of the $glass/air$ interface is higher than the total surface tension of the interfaces $wine/air$ and $wine/glass$.
While the film rises along the glass wall the alcohol in the film evaporates.
Alcohol lowers the surface tension of water, and therefore, 
the surface tension at the top of the film is higher than at the bottom.
The Marangoni flow drives then the fluid up the film until gravity forces the excess fluid to drop back in tears.

%%%%%%%%%%%%%%%%%%%%%%%%%%%%%%%%%%%%%%%%%%%%%%%%%
\subsection*{Young-Laplace equation}
The last equation governing the physics of interfaces is the Young-Laplace equation.
Lets imagine that a bubble of area $A$ and volume $V$ is deformed 
and lets denote $P_{out}$ the external bulk fluid pressure and 
$P_{in}$ the internal bulk fluid pressure. 
The total amount of bulk plus boundary work 
is then given by 
\be
\delta W=  (P_{out}-P_{int})\delta V+ \gamma \delta A~. \nn
\ee
The condition of mechanical equilibrium then leads to the Young-Laplace equation: 
\be\label{LY}
\Delta P + \gamma \theta_S =0~,
\ee
where $\Delta P = (P_{out}-P_{int})$ and 
\be
\theta_S \equiv \left(\frac{\partial A}{\partial V}\right) 
\ee
is the bubble curvature.
If the bubble is spherical with radius $R$ it is given by $\theta_S = \frac2R$.
If not it can be written in terms of the two principal radii of curvature $R_{1},R_{2}$ as the mean curvature 
\be
\theta_S =\f{1}{R_1}+\f{1}{R_2}.
\ee
It can also be expressed as the divergence of the normal vector to the surface.
If $\s$ is a unit outward vector normal to the surface then
\be\lb{b_theta}
\theta_S= \nabla_{i} s^{i}.
\ee
The Young-Laplace equation \Ref{LY} determines the mechanical-equilibrium shape of interfaces.
It means that the curvature of the surface creates an extra normal force $\g \theta$ directed inside the bubble center. 
This force is responsible for the capillarity rise in small tubes, 
the buoyancy of small insect on water, the capillary suction of parallel plates 
and the Rayleigh instability of cylindrical flow, among other things \cite{dG}.
 
%%%%%%%%%%%%%%%%%%%%%%%%%%%%%%%%%
\subsection*{Force law}
\begin{figure}[h]
 \begin{center}
\includegraphics*[scale=0.16]{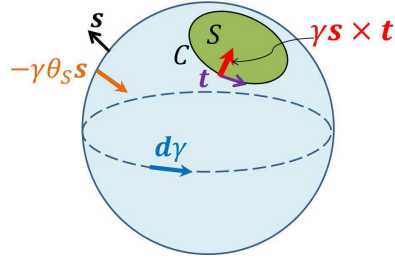}
 \caption{Surface tension as a force per unit length}
 \label{fig:g_force}
 \end{center}
 \end{figure}

The above discussions showed that the presence of a surface tension leads to 
both an inward normal force $- (\gamma \theta_S) \s$ and a tangential force $\bm{\rd} \gamma$.
Note here that $\brd$ represents tangential derivative, which is defined by $\rd_A f\equiv q_A{}^i\N_i f$, 
where $q_{ij}$ is 2d metric for the interface such that $\d_{ij}=s_is_j+q_{ij}$. 
The application of Stokes theorem to an element of surface $S$ with boundary $C=\partial S$ gives 
the total force on $S$ due to surface tension: 
\be
\bm{F}_{S}=\int_{S} (\bm{\rd} \gamma- (\gamma \theta_S) \s ) \rd A
= \int_{C} \g (\s\times \t)  \rd \ell~,
\ee
where $\t$ is the unit vector tangent to $C$, $\times$ denotes the cross product and $\rd \ell$ the line element (see fig.\ref{fig:g_force}).
This shows that the {\it surface tension is the force per unit length tangential to the surface S 
but normal to the boundary curve C and directed inwardly}. 
Indeed, in the 2d system, dimensionally $[\g]=(force)\times (length)^{-1}$.

%%%%%%%%%%%%%%%%%%%%
\subsection*{Wetting}
The interpretation of surface tension as a force per unit length acting on the contact line of interfaces explains the phenomena of wetting.
Suppose a drop of a liquid placed at  a  solid/gas interface,
and lets denote respectively by  $\g$, $\g_{SL }$, and $\g_{SG}$ 
the gas/liquid, solid/liquid and solid/gas  
 interface tensions.
 Let's consider 
\begin{equation}
S=\g_{SG}-(\g_{SL}+\g)~.\nn
\end{equation}
\begin{figure}[h]
 \begin{center}
\includegraphics*[scale=0.2]{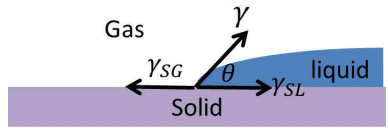}
 \caption{Wetting process}
 \label{fig:contact}
 \end{center}
 \end{figure}
%Since the interface tension is a force per unit length of the contact line, 
If $S>0$, the drop will spread out, that is, wetting will occur.
If $S<0$, wetting happens partially, and an equilibrium angle holds, which is determined by 
the law of Young-Dupre: 
\begin{equation}
\g \cos \theta = \g_{SG } - \g_{SL}~. \nn
\end{equation}

%%%%%%%%%%%%%%%%%%%%%%%%%%%%%%%%%%%%%%%%%%%
\subsection*{Dynamical Young-Laplace equation: conservation law of normal momentum}
In the dynamical case the Young-Laplace equilibrium condition \Ref{LY} is not always satisfied.
If $\Delta P +\gamma \theta_S >0~(<0)$ there is a force directed inward (outward). %the inside of the bubble, . 
This force  results in a mass transfer. 
This happens for instance in the case of evaporation or condensation, 
where matter is transported across the surface. 

By treating appropriately the interface as the thin limit of a finite size layer, 
we can derive the equation governing the dynamics of the interface:
 \be\label{DLY}
 \rd_t  \bar\pi_{r}+ (\rd_{A} v^{A}) \bar{\pi}_{r} = -(\Delta P +\gamma \theta_S),
 \ee
which is derived in Appendix.
Here $\bar \pi_r $ is the bubble radial momentum density.

%%%%%%%%%%%%%%%%%%%%%%%%%%%%%%%%%%%%%%%%
\subsection*{Generalization to viscous bubbles}
The previous equations (\ref{interfU},\ref{marangoni},\ref{DLY}) express the conservation laws for a non-viscous bubble.
In the case where the bubble is viscous, we have to add to these laws the influence of the 2d viscous stress tensor $\bm\Theta$, 2d rate of strain tensor $\bm\Sigma$ and 2d heat flow vector $\bm{q}$. 
These tensors enter into the conservations of tangential momenta and interfacial internal energy 
in exactly the same way that bulk viscous stress and bulk rate of strain enter into the conservations of energy and momenta as described in the previous section.
In this context an external heat transfer analogous to  $\dot Q_{\mathrm{rad}}$ also enters naturally since there can be heat crossing the interface along the normal direction.
Note finally that the conservation law of normal momentum \Ref{DLY} is modified non-trivially.
The general form of this modification for a viscous bubble is not known in the thermodynamical literature, and thus  our approach gives a definite proposal for what this can be.
In the next section we will analyze gravity equations
and show that they correspond to a viscous generalization   of the thermodynamic laws of interfaces. 

%%%%%%%%%%%%%%%%%%%%%%%%%%%%%%%%%%%%%%%%%%%%%%%%%%%%%%%%%%%%%%%%%%%%%%%%%%%%%%%%%%%%%%%%%%%%%%%%%%%%%
\section{Gravity equations for the screen}
We now change gears and study the gravity equations 
for a timelike membrane which appears as the time evolution of  the \textit{gravitational screen} \ci{L}.

%%%%%%%%%%%%%%
\subsection{Screen energy-momentum tensor}
\begin{figure}[h]
 \begin{center}
\includegraphics*[scale=0.2]{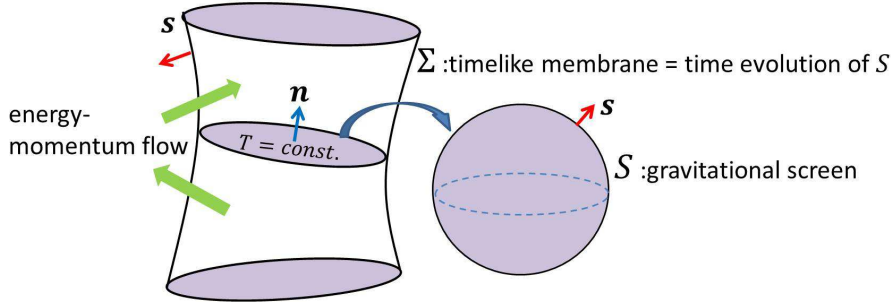}
 \caption{A gravitational screen and vectors}
 \label{fig:screen}
 \end{center}
 \end{figure}

We now consider a 3d timelike hypersurface $\Sigma= S\times {R}$, where $S$ is a $2$-sphere, embedded in 4d space-time with unit normal $\s$ satisfying $\s^{2}=1$. 
$\s$ is chosen  to point toward the outside of the region screened by the membrane (see fig.\ref{fig:screen}). 
We denote by ${h}_{ab}\equiv g_{ab}- s_{a}s_{b}$ the induced metric on membrane $\Sigma$ and by
$H_{ab} \equiv {h}_{a}{}^{a'}{h}_{b}{}^{b'} \nabla_{a'}s_{b'}=h_{a}{}^{a'}h_{b}{}^{b'} \frac12 \cL_{\s}h_{a'b'}$, 
the extrinsic curvature tensor of the membrane.
Here $g_{ab}$ is the 4d space-time metric. 
Note that this membrane can be chosen arbitrarily as we like.

Such a screen can be considered to carry a \textit{screen energy-momentum tensor} $S_{ab}$ given by
\be\lb{SEM}
S_{ab} \equiv \frac{1}{8\pi G} \bar S_{ab}=\frac{1}{8\pi G} ( H h_{ab} - H_{ab} )~, 
\ee
where $H=h^{ab}H_{ab}$ and $\bar S_{ab}\equiv H h_{ab} - H_{ab}$.
This is justified by using the Israel's junction condition for a timelike surface \ci{Israel, Poisson} 
and is similar in spirit to what is done in the black-hole membrane paradigm \ci{Price, Membrane}.

Indeed, it is well-known that if we consider  a space-time obtained by gluing two half space-time $(M_{\pm},g_{\pm})$ along a timelike hypersurface $\Sigma$, 
we have to demand that  both induced metrics agree on $\Sigma$, in other words  $h_{ab}^{+}=h_{ab}^{-}$, 
in order to be able to define the Riemann tensor. 
The Riemann tensor can contain a distributional contribution supported only on $\Sigma$.
The Weyl component of this contribution vanishes, 
while the Ricci component is characterized by the discontinuity of the extrinsic curvature $H_{ab}$ across $\Sigma$. 
Thus, the Einstein tensor is described as 
\bea
 G_{ab} &=& G_{ab}^{+}\Theta_{\Sigma}^{+}(x) +G_{ab}^{-}\Theta_{\Sigma}^{-}(x)+ \bar S_{ab} \delta_{\Sigma}(x)\\
 \bar S_{ab}&=&[H]^+_- h_{ab}-[H_{ab}]^+_-
\eea
where $\Theta^{\pm}_\Sigma$ are  distributions supported on the each half space and 
$\delta_{\Sigma}$ is a characteristic distribution of $\Sigma$: $\int \sqrt{|g|} f \delta_{\Sigma}= \int_{\Sigma}\sqrt{h} f$. 
Using the Einstein equation, the distributional contribution of $G_{ab}$ can be considered to come from 
the presence of the surface energy-momentum tensor associated with the timelike membrane $\Sigma$, $S_{ab}=\f{1}{8\pi G}\bar S_{ab}$. 

Next let's employ  the membrane paradigm idea.
In reality, there is no such a surface energy-momentum tensor on $\Sigma$ since in our case  we are here just postulating a timelike surface in a space-time. 
However, in the membrane paradigm,
 the space-time is supposed to be terminated by the timelike surface.
 This leads to an holographic point of view where the interior of the membrane 
 is entirely replaced by the presence of a physical membrane.
 In this case  $H_{ab}|_-=0$ and the membrane energy-momentum tensor is given by $S_{ab}$.
 The remarkable point is that the replacement of the interior by a pure boundary does not affect the physics outside $\Sigma$. The boundary physics actually  replaces the actual inside spacetime as long as the screen is assigned a non vanishing energy momentum tensor $S_{ab}$.
 
 The same energy momentum tensor appears in Brown-York \cite{B-Y} analysis and is often referred as 
the Brown-York energy-momentum tensor.
In this point of view it comes from the boundary variation of the Hamilton-Jacobi functional as 
\be
S_{ab} = -\frac{2}{\sqrt{|h|}} \frac{\delta S_{\mathrm{H.J}}}{\delta h^{ab} },
\ee
which is a relativistic generalisation of the usual definition of energy from the Hamilton-Jacobi functional $E =-\pa S/\pa t$.
The membrane paradigm interpretation is more adapted to our analysis and was performed earlier so we give it precedence.

We can use the screen energy-momentum tensor \Ref{SEM} to rewrite the gravity equations in an interesting form.
Using the Codazzi \Ref{Codazzi} and the Gauss equations \Ref{Gauss} in Appendix, 
the Einstein equations $G_{ab}=(8\pi G) T_{ab}$ projected on the timelike brane are expressed as 
\bea\lb{conserve_eq}
D_bS^{ba} &=&- T_{\s b}h^{ba}, \\
\lb{constraint_eq}
R(h)+H^{ab}H_{ab}-H^2&=&-(16\pi G)T_{\s\s}~,
\eea
where $D_{a} V_{b} = h_{a}^{a'}h_{b}^{b'}\nabla_{a'}V_{b'}$ denotes the covariant derivative on the timelike membrane and $T_{\s a}\equiv T_{ba}s^{b}$. 
The first equations express the screen energy-momentum conservation laws under an external source. $-h^{ab}T_{b\s}$ 
represents the  radial inflow across the surface of the $a$-component  energy-momentum.
In the membrane point of view is interpreted as the energy-momentum in-flow from external forces and radiation.
The second equation corresponds to a constraint equation for ``radial direction", 
which is analogous to the constraint equation of general relativity for time foliation.
It controls the amount of radial momentum flowing across the screen.

%%%%%%%%%%%%%%%%%%%%%%%%%%%%%%%%%%%%%%%%%%%%%%%%%%%%%%%%%
\subsection{$2+2$ decomposition}
The screen evolution $\Sigma$ is a timelike 3d submanifold embedded in the 4d spacetime manifold.
In order to understand its dynamics we introduce on it a space-time decomposition \ci{L}.
We assume that there is a slicing of $\Sigma $ such that the constant time slices are 2d spacelike surfaces $S$
which are topologically 2d spheres and the timelike membrane is viewed as the screen evolution.
We denote by $\n$ the unit timelike normal to the slices (i.e. $\n^2=-1$) and by $\rho$ the screen lapse function (see fig.\ref{fig:screen}).
That is, we have that
\be
\n \hat{=}-\rho \brd T,
\ee
where $T$ is the time function and the equality $\hat =$ is valid on the screen evolution.
The time lapse $\rho$ plays the role of the Newtonian potential associated with the membrane's foliation, as we will see later.
Note that by construction $\n \! \cdot \! \bm{s}=0$ holds, and 
the two dimensional metric $q_{AB}$ characterizes the intrinsic geometry of $S$.
Here the index $A,B...$ represents components tangent to $S$, and 
in the following we will denote by $\rd_{A}$ the covariant derivative on $S$, 
which is defined as $\rd_A v^B\equiv q_A{}^a q^B{}_b \N_a v^b$. 
$q_{ab}$ is related to the screen metric by $h_{ab}= q_{ab}-n_{a}n_{b}$, and thus, we have 
\begin{equation}\lb{g}
g_{ab}=-n_an_b + s_a s_b + q_{ab}~,
\end{equation}
which will be used to decompose geometric quantities in the following.

The extrinsic geometry of $S$ is first characterized by 
the {\it extrinsic curvature tensors}
\be
\Theta_{\bm{n}AB} \equiv q_{A}{}^{a}q_{B}{}^{b}\nabla_{a} n_{b},\qquad 
\Theta_{\bm{s}AB} \equiv q_{A}{}^{a}q_{B}{}^{b}\nabla_{a} s_{b},
\ee
whose trace part is denoted by 
\be
\theta_{\bm{\ell}}= q^{AB}\Theta_{\bm{\ell}AB},
\ee
for  a normal vector $\bm{\ell}$.
The extrinsic geometry is  also characterized by the {\it normal one-form}: 
\be
\omega_{A}\equiv q_{A}{}^{a} (\s\!\cdot \nabla_{a}\n).
\ee
Thus, the extrinsic geometry is governed by $(\Theta_{\bm{n}AB},~\Theta_{\bm{s}AB},~\omega_{A})$.

Note here that the screen evolution $\Sigma$ we are considering is embedded into 4d spacetime and its normal is $\s$. Its  
 intrinsic metric is $h_{ab}$. The screen itself $S$ is embedded inside $\Sigma$ with normal $\n$. From the 4d point of view the projection onto 
 the directions tangential to $S$ is done by $q_{ab}$.
 The projection along the directions normal to $S$ is done by  $N_{ab}\equiv -n_{a}n_{b}+ s_{a}s_{b}$. 
 Here we distinguish the ``intrinsic geometry'' of $S$ encoded in $q_{ab}$ from its ``extrinsic geometry'' encoded in the normal derivatives of $q$ 
 which is distinct from its ``normal geometry'' encoding the normal derivatives of the normals\footnote{This denomination does not  assume 
 that the normal to $S$ is integrable just that there are geometrical data that depends purely on the normals.}.
 The ``normal geometry" , is encoded into three types of data. 
The first ones are the {\it tangential accelerations}:
\be
a_{\n}^{A}\equiv q^{A}{}_{a}(\nabla_{\n}\n)^a,~~~a_{\s}^A\equiv- q^{A}{}_{a} (\nabla_{\s}\s)^a,~~~a_{\n\s}^{A}\equiv \frac12q^{A}{}_{a} (\nabla_{\s}\n + \nabla_{\n}\s)^{a},
\ee
which measure how the normal deformations varies on the sphere $S$.
%For instance we can express $a_{\n}$ in terms of the lapse vector as $ a_{\n A}= \rd_{A}\phi/\phi$.
The second one is the {\it twist vector}:
\be
j^{A}\equiv q^{A}{}_{a} [\n,\s]^{a},
\ee
which measures the lack of integrability of the normal planes.
And the final ones are the \textit{normal accelerations}:
\be
\gamma_{\n}\equiv \s\!\cdot \nabla_{\n}\n,\qquad 
\gamma_{\s}\equiv -\n\!\cdot \nabla_{\s}\s.
\ee
which measure how the normal varies away from the sphere.
Note that $\gamma_{\n}$ is the radial acceleration of an observer flowing along the screen.
It represents the total acceleration of an observer following a screen geodesic.

The above geometric quantities decompose the 4d spacetime into the 2d surface $S$ and the 2d normal surface $T^{\perp}S$, 
which we will call \textit{2+2 decomposition}. 
We can now use this decomposition to project the conservation laws \Ref{conserve_eq} onto timelike and spatial directions: 
 \be\lb{conserve_eq2}
 (D_{b} S^{b}{}_{a})n^{a} = - T_{\s \n},\qquad (D_b S^b{}_a) q^a{}_A=- T_{\s A},
 \ee 
which can be interpreted as the conservation of energy and momenta of the timelike screen, respectively. 
Indeed $T_{\s \n}$ represents the amount of matter energy flowing out the screen 
while $T_{\s A}$ represents the amount of matter momenta flowing out the screen.

The screen energy-momentum tensor $S_{ab}$ can be decomposed into 
energy density ${\epsilon}\equiv S_{\n\n}$, momentum density $\pi_{A} \equiv - q_{A}{}^{a}S_{a\n}$ 
and total pressure tensor $\Pi_{AB}\equiv q_{A}{}^{a} q_{B}{}^{b} S_{ab}$ as
\begin{equation}
S_{ab}=\e n_a n_b + \pi_a n_b + \pi_b n_a +\Pi_{ab}~.
\end{equation}
These components can be explicitly expressed in terms of the 2+2 decomposition: 
\begin{equation}\lb{H}
H_{ab}=-\g_{\n}n_a n_b +\omega_a n_b + \omega_b n_a +\Theta_{\s ab},\quad H=\g_{\n}+\theta_{\s}~.
\end{equation}
We eventually find that 
\be\lb{S_exp}
8\pi G S_{ab}=\bar S_{ab} = -\theta_{\s} n_{a}n_{b} - \omega_{a} n_{b} - \omega_{b}n_{a} + (  \tilde{\Theta}_{\s ab} + \gamma_{\n} q_{ab}  )~,
\ee
where we have denoted by $\tilde{}$ the operation
 \be
 \tilde{\Theta}^{AB}\equiv \theta q^{AB}-\Theta^{AB} .
 \ee
 This suggests already the identification of energy, momenta and pressure tensor to be given by
\be
\epsilon=  -\left(\frac{\theta_{\s}}{8\pi G}\right), \qquad
\bm\pi = -\left(\frac{\bom}{8\pi G}\right), \qquad 
 \bm{\Pi}=
 \left(\frac{\gamma_{\n}}{8\pi G}\right)\bm{q} +
 \frac{\bm{\tilde{\Theta}_{\s}}}{8\pi G}.
\ee
This interpretation will be confirmed later up to a rescaling. 
In order to see this we now look at the equations this tensor satisfies.

%%%%%%%%%%%%%%%%%%%%%%%%%%%%%%%%%%
\subsection{ Conservation of screen momenta}
We start by analyzing 
the  conservation equation of the screen momenta in \Ref{conserve_eq2}: $-8 \pi G T_{\s A}= (D_{b} \bar S^{b}{}_{a})q^{a}{}_{A}$.
We expand each term in this equation by using the decomposition \Ref{S_exp} and 
making  use of the identity 
\be\label{da}
D_{a}V^{a}= \rd_{A} V^{A} + a_{\n A} V^{A} 
\ee
for a vector  tangent to $S$, i-e such that $q^a{}_bV^{b}=V^a$. 
\bea
-(8 \pi G)T_{\s A} 
&=& (D_{b}(\gamma_{\n}q^{b}{}_{a} + \tilde{\Theta}_{\s a}^{b})
 - D_{b}(n^{b}\omega_{a}) - D_{b}(\om^{b}n_{a}) - D_{b}(n^{b}\theta_{\s} n_{a}) ) q^{a}{}_{A} \nn\\
&=& (\rd_{A} + a_{\n A}) \gamma_{\n} + (\rd_{B}+a_{\n B}) \tilde{\Theta}_{\s A}^{B} -q_{A}{}^{a}\N_{\n}\om_{a} - \theta_{\n} \om_{A} - \om^{B} \Theta_{\n BA}  
-\theta_{\s}\ba_{\n A} \nn \\
&=& -{\cL}_{\n}\om_{A}- \theta_{\n} \om_{A}  +(\rd_{A} + a_{\n A}) \gamma_{\n}+ (\rd_{B} + a_{\n B}) \tilde{\Theta}_{\s A}^{B}  -\theta_{\s} a_{\n A}~,\nn
\eea
where $\cL$ denotes the Lie derivative and we have used that 
\be
\cL_{\n}\om_{A} = \nabla_{\n}\om_A + \om_B \Theta_{\n}^B{}_A.\nn
\ee
We can therefore write this conservation equation as 
\be
\boxed{
{\cL}_{\n}\bom+ \theta_{\n} \bom =  (8\pi G)  T_{\s \cdot}+
(\brd + \ba_{\n}) \gamma_{\n} +  
(\brd + \ba_{\n })\dd  \bm{\tilde{\Theta}_{\s } } 
-\theta_{\s} \ba_{\n }
}~,\label{Tsa}
\ee
where $T_{\s \cdot}$ represents the 2d-vector momentum from the outside with components $T_{\s A}$.
%%%%%%%%%%%%%%%%%%%%%%%%%%%%%%%%%%%%
\subsection{Conservation of screen energy}
We now focus on the  equation expressing the  conservation of the screen energy in \Ref{conserve_eq2}:$-(8 \pi G)T_{\s \n}= (D_{b} \bar S^{b}{}_{a})n^{a}$.
Expanding each term and using \Ref{da} in the same way, we get
\bea
-(8 \pi G)T_{\s \n} 
&=& \left[D_{b} (q^{b}{}_{a}\gamma_{\n})+ D_{b}{\tilde{\Theta}}^{b}_{\s a} 
- D_{b}(n^{b}\om_{a}) - D_{b}(\om^{b}n_{a}) - D_{b}(n^{b}\theta_{\s} n^{a}) \right] n_{a}\nn\\
&=& -\gamma_{\n}\theta_{\n}- {\tilde{\Theta}}^{B}_{\s A}(D_{B}n^{A}) -(\N_{\n}\om_{a})n^{a}+ D_{b}\om^{b} + D_{b}(n^{b}\theta_{\s}) \nn \\
&=& -\gamma_{\n}\theta_{\n}- {\tilde{\Theta}}^{B}_{\s A} \Theta_{\n B}^{A} + \om_{A} a_{\n}^{A}+ (\rd_{A} + a_{\n A})\om^{A} + \theta_{\n}\theta_{\s} + \N_{\n}\theta_{\s},\nn
\eea
In summary the conservation of screen energy reads
\be
\boxed{
(\cL_{\n} + \theta_{\n})\theta_{\s}
=-(8 \pi G)T_{\s \n}+
\gamma_{\n}\theta_{\n} + \bm{{\tilde{\Theta}}}_{\s}:{\bm{\Theta}_{\n}}
- (\brd + 2 \ba_{\n})\dd \bom
}~.\label{Tsn}
\ee
%%%%%%%%%%%%%%%%%%%%%%%%%%%%%%%%%%%%%%%%%%%%%%%%%%%%%%%%%%%%%%%%%%%%%%%%%%%

%%%%%%%%%%%%%%%%%%%%%%%%%%%%%%%%%%%%%%%%%%%%%%%%%%%%%%%%%%%%%%%%%%%%%%%%
\subsection{Radial Constraint equation}
We finally analyze the decomposition of the radial constraint equation \Ref{constraint_eq}: 
$(16 \pi G)  T_{\s\s}=  H^{2} -  H_{ab}H^{ab} - R(h)$.
Using \Ref{H} we evaluate the quadratic terms as
\begin{equation}
H^2-H_{ab}H^{ab}=\bm{\tilde \Theta}_{\s}:\bm{\Theta}_{\s} + 2 \bm{\omega}^{2}  +2 \gamma_{\n}\theta_{\s}~. \nn
\end{equation}
%\footnote{ It is interesting to note that we have for a 2dimensional matrix, the identity$$ \det{\bm{\Theta}} = \f12   \bm{\tilde{\Theta}}:\bm\Theta.$$}
The Ricci equation \Ref{Ricci}, controls the way the 3-dimensional and 2-dimensional curvatures are related:
\bea
R(h)
&=& R(q) + \bm{\Theta}_{\n }:\bm\Theta_{\n}- \theta_{\n}^{2}
 + 2 D_{a}(n^{a} \theta_{\n}-a^{a}_{\n})\nonumber
\\
& =& R(q) - \bm{\tilde\Theta}_{\n }:\bm\Theta_{\n}
+ 2( \N_{\n} +\theta_{\n})\theta_{\n} - 2(\bm\rd +\bm{a}_{\n})\dd \ba_{\n}  \nn.
\eea
Thus, we arrive at 
\begin{equation}
\label{Tss}
\boxed{
( \cL_{\n} +\theta_{\n})\theta_{\n} 
= -(8 \pi G)  T_{\s\s} +\gamma_{\n} \theta_{\s}  
+\bm{\om}^{2} + \f12 \bm{\tilde{\Theta}}_{\n }:\bm\Theta_{\n}
 + \f12 \bm{\tilde \Theta}_{\s}:\bm{\Theta}_{\s}  
 - \f12 R(q) + (\bm\rd +\bm{a}_{\n})\dd \ba_{\n}}~.
\end{equation}

%%%%%%%%%%%%%%%%%%%%%%%%%%%%%%%%%%%%%%%%%%%%%%%%%%%%%%%%%%%%%%%%%%%%%%%%%%%%%%%%%%%%%%%%%%%%%%%%%%
\section{The gravity-thermodynamics dictionary}
We now show that the previous equations reduce to the thermodynamic equations for a viscous bubble.
First, we introduce a time flow vector $\t$ normal to the sphere $S$, 
and we denote its norm by 
\be
\t^{2}= - \rho^2.
\ee
In the following we also denote by $\ts$ the dual vector orthogonal to 
$\t$ having the  norm $ \ts\dd \ts= \rho^2$ and preserving the orientation of the normal planes.
If $(\n,\s)$ denotes a basis of $T^{\perp}S$ this means that we can write 
\be
\t=\rho(\cosh\beta \n + \sinh\beta \s),\qquad
\ts= \rho(\cosh\beta \s + \sinh\beta \n).
\ee
The boost angle $\beta$ corresponds to a choice of frame of the normal planes.
The condition that $\t$ is the vector generating the time flow on $\Sigma = S\times{R}$ requires
\begin{equation}\lb{t_cond}
q_A{}^a\cL_{\t}t_a=0~.
\end{equation}
Using this we can now rewrite the previous conservation 
equations in terms of $(\t,\ts)$,
which is more natural since time evolution is observed not by a basis $\n$ but by a time vector $\t$. 
First lets assume that we are in the frame where $\beta=0$, so that 
\begin{equation}\lb{t_rescaled}
\t =\rho \n,~~\ts=\rho \s~.
\end{equation}
Then, the condition \Ref{t_cond} leads to an expression of the acceleration $\bm a_{\n}$: 
\begin{equation}\lb{a_condition}
\bm a_{\n} = \brd \ln \rho = \brd \phi_N~,
\end{equation}
where $\phi_N=\ln \rho$ is the Newtonian potential associated to the screen. 
Next we introduce new quantities by rescaling as 
\begin{equation}\lb{rescaled}
\g_{\t}\equiv \rho \g_{\n},~~\bm{\Theta}_{\ts}\equiv\rho\bm{\Theta}_{\s},~~\bm{\Theta}_{\t}\equiv \rho\bm{\Theta}_{\n}~.
\end{equation}
If one  multiplies (\ref{Tsa}) by ${\rho}$, we can express the gravity equation of momentum conservation as:
\be
\boxed{
D_{\t}\bom+ \theta_{\t} \bom = 
\brd  \gamma_{\t} +  
\brd \dd  \bm{\tilde{\Theta}_{\ts } } 
-\theta_{\ts} \brd \phi_N + (8\pi G)  T_{\ts A}
}~,\label{Tsat}
\ee
where by $D_{\t}$ we have denoted
$
D_{\t} v^{A} \equiv q^{A}_{a}\cL_{\t} v^{a} .
$
Similarly by multiplying (\ref{Tsn}) by ${\rho}^{2}$ we obtain that
\be
\boxed{
(D_{\t} + \theta_{\t})\theta_{\ts}
=
\gamma_{\t}\theta_{\t} + \bm{{\tilde{\Theta}}}_{\ts}:{\bm{\Theta}_{\t}}
- \brd \dd(\rho^{2} \bom)
+\theta_{\ts}D_{\t}\phi_N
 -(8 \pi G)T_{\t \ts}}~.\label{Tsnt}
\ee

In order to interpret these equations physically, let's go back to the thermodynamic equations.
First, combining the local first law \Ref{1st} and the equation of the Newton potential \Ref{dphi} 
we obtain  a ``generalized" first law: 
\begin{equation}\lb{u+phi}
(\rd_t + \sigma)(u+\rho \phi)=-p\sigma + \bm \Theta : \bm \Sigma -\brd \cdot \bm q + \rho \rd_t \phi+ \dot Q_{\mathrm{rad}}~,
\end{equation}
which is the equation of time evolution for the sum of internal energy density and gravitational potential density, $u+\rho \phi$.
Note here that when usual thermodynamic equations are written down, we will use the notations introduced in the section of thermodynamics.
What  is now clear and quite remarkable is that,
after   dividing them by $-(8\pi G)$, the equations
\Ref{Tsat} and \Ref{Tsnt} take exactly  the same forms as 
the thermodynamic equations \Ref{momenta} and \Ref{u+phi},
which express the conservation of momenta, 
and the sum of internal energy and gravitational energy of a viscous bubble, respectively!
This is true if one identifies the projected Lie derivative $D_{\t}$ with the particular derivative $\rd_{t}$.
This  provides  a clean correspondence between the fluid dynamics and gravity that we now spell out.

Let us first recall that
$-T_{\s \t}$ represents matter energy flow density going in the direction $\s$ measured by an observer $\t$, 
and $T_{\s A}$ represents $A$-component of matter momentum flow density going in the direction $\s$
\footnote{
Indeed, the total momenta and energy associated with a slice 
$\bar{\Sigma}$ with induced metric ${\bar{h}}$ is 
\be
P_{A}= -
\int_{\bar\Sigma} T_{A \n} \sqrt{\bar{h}},\qquad E=\int_{\bar\Sigma} T_{\t \n} \sqrt{\bar{h}}\nn
\ee
Suppose that $\t$ is a killing vector, then $T^{a}{}_{\t}$ is a conserved tensor and  using the Gauss law 
on a timelike cylinder with space like boundaries $\Sigma_{\pm}$ and timelike boundary $S\times I$ where $I$ is a time interval, by the Gauss Law we have that
\be
0
%= \int_{V}\pa_{a}T^{a}{}_{\t}\sqrt{|g|}
=-\int^{\bar{\Sigma}_+}_{\bar{\Sigma}_-} T_{\n\t}  \sqrt{\bar{h}}+ \int_{S\times I}T_{\s\t}  \sqrt{|{h}|}=0,\quad\mathrm{thus}\quad \Delta{E}= \int_{S\times I}T_{\t\s} \sqrt{|{h}|}.\nn
\ee
}.
Thus, $T_{\ts \t}$ and $-T_{\ts A}$ represents the matter energy and momentum flow density 
entering into the screen, redshifted by the lapse factor $\rho$
\footnote{A similar interpretation is discussed in the original membrane paradigm \ci{Price, Membrane}, 
although a null limit is taken there.}.

If we assume that $\t$  represents the relativistic velocity of the fluid, which could be checked by reconsidering this formulation in a framework of relativistic hydrodynamics \ci{LY2}, 
it is natural to identify  the projected Lie derivative $D_{\t}$ along $\t$
as the Lagrange derivative $\rd_t$. 
This means  that what plays the role of the compressibility coefficient $\sigma=\p_iv^i$
is given by the temporal expansion $\theta_{\t}=D_{\t}\ln \sqrt{q}$, 
which measures how the size of the 2d surface evolves along $\t$. 
This implies that the 2d measure density $\sqrt{q}$ corresponds to the specific volume  $v=1/\rho$.

Here let us start with the analysis of the momenta conservation (\ref{Tsat}). 
We see that the fluid momenta can be identified with $-\frac{\bom}{8\pi G}$. 
%As we have seen in the above, $-T_{\ts A}$ represents the flow of matter momenta $A$ in the direction $-\ts$. 
From the fluid perspective $-T_{\ts A}$ represents the external forces $f_{A}$ that matter imprints on the bubble fluid.
The first term in the RHS of \Ref{Tsat} is a Marangoni force term, 
which tells us that $-\frac{\gamma_{\t}}{8\pi G}$ plays the role of the surface tension $\gamma$, 
or the minus pressure $-p_{2d}$, since this system corresponds to a 2d fluid, so \Ref{g=-p} holds.  
Also $-\frac{\bm{\tilde\Theta}_{\ts}}{8\pi G}$ plays the role of the viscous stress tensor $\bm{\Theta}$.
Finally the third term in the RHS of \Ref{Tsat} appears to be equivalent to a force term from the Newtonian gravity 
where $\phi_N$ \textit{is} the Newtonian potential 
while $-\frac{\theta_{\ts}}{8\pi G}$ represents the Newtonian mass density.
Thus, \Ref{Tsat}, which is originally the equation of the momentum conservation for the bulk, represents 
the law of momentum conservation of a 2d viscous bubble under the gravitational and external forces. 

This interpretation can also be confirmed consistently in the energy conservation equation \Ref{Tsnt}.
Indeed the surface tension $\gamma=-\frac{\gamma_{\t}}{8\pi G}$ appears as a work term $-\frac{\gamma_{\t}}{8\pi G}\theta_{\t} $.
As we have mentioned, 
if $\t$ corresponds to the fluid velocity, 
it is natural to interpret $\bm\Theta_{\t}$ as the rate of strain tensor $\bm\Sigma=\bm \p \v$.
According to this and the interpretation of $-\frac{\bm{\tilde\Theta}_{\ts}}{8\pi G}$ as the viscous stress tensor in \Ref{Tsat}, 
the term $\frac{-\bm{\tilde\Theta}_{\ts}}{8\pi G}:\bm \Theta_{\t}$ appears consistently as the viscous dissipation term. 
This dissipative term corresponds to a measure of the gravitational waves entering the screen, 
since in the null limit of the screen this quantity is related to Weyl tensor \ci{Hartle, Ashtekar2, Chirco}. 
The divergence term represents the heat flux $ \bm{q}= -\f{\rho^2 \bm \om}{8\pi G}$, 
and $T_{\ts \t}$, which is the matter energy flux density entering to the screen, corresponds to the radiative heat production $\dot Q_{\mathrm{rad}}$. 
Furthermore, $-\frac{\theta_{\ts}}{8\pi G}$ appears again as the Newtonian mass in the time-dependent Newtonian potential term $\rho \rd_t \phi$. 
Finally, in order to compare \Ref{Tsnt} with \Ref{u+phi}, 
we have to identified $-\frac{\theta_{\ts}}{8\pi G}$ as the sum of the internal energy density and the gravitational potential energy, $u+\rho \phi$. 
Thus, we can understand \Ref{Tsnt} as the law of 
the ``generalized" first law of thermodynamics for a 2d viscous bubble in the manner consistent with \Ref{Tsat}. 

One peculiarity of the above analysis is that 
the Newtonian mass is identified with the sum of internal and gravitational energy,
and that the heat flux is in the direction of the momenta.
However, this can come from the fact that this system is relativistic, 
while it is not possible to distinguish mass, internal, and gravitational energy in a relativistic fluid \ci{Landau_F}, 
which we can also check by discussing this formulation as a relativistic fluid \ci{LY2}. 
%\footnote{We can also discuss this formulation as a relativistic fluid \ci{LY2}.}.
The dictionary between bubble fluid and gravity is presented in  table \ref{GF1}.
\begin{center}
\begin{table}[ht]
\caption{Dictionary 1 for gravity-bubble thermodynamics correspondence}
\centering
\begin{tabular}{r c c c l}
\hline\hline
 Thermodynamical name  &   &  symbols &  & Gravity name \\ [0.5ex] % inserts table %heading
\hline\\
specific volume:&$v=1/\rho$ &$\leftrightarrow$ &${\sqrt{q}}$ &
:2d measure \\
\\
compressibility coefficient:&$\sigma$ &$\leftrightarrow$ &$\theta_{\t}$ &
:expansion \\
\\
\raisebox{2.5ex}{internal + gravitational energy:}&\raisebox{2.5ex}{$u+\rho \phi $} &\raisebox{2.5ex}{$\leftrightarrow$} &\raisebox{2.5ex}{$\displaystyle-\frac{\theta_{\ts}}{8\pi G}$} &
\raisebox{2.5ex}{:inward radial expansion} \\
surface tension:&$\gamma $ &$\leftrightarrow$ &$\displaystyle-\frac{\gamma_{\t}}{8\pi G}$ &
:inward radial acceleration \\
\\
2d pressure:&$p_{2d} $ &$\leftrightarrow$ &$\displaystyle \frac{\gamma_{\t}}{8\pi G}$ &
:outward radial acceleration \\
\\
\raisebox{2.5ex}{ Newtonian mass density:}&\raisebox{2.5ex}{$\rho $} &\raisebox{2.5ex}{$\leftrightarrow$} &\raisebox{2.5ex}{$\displaystyle-\frac{\theta_{\ts}}{8\pi G}$}&
\raisebox{2.5ex}{:inward radial expansion} \\
\raisebox{2.5ex}{ Newtonian potential:}&\raisebox{2.5ex}{$\phi $} &\raisebox{2.5ex}{$\leftrightarrow$} &\raisebox{2.5ex}{$\phi_N=\ln \rho$}&
\raisebox{2.5ex}{:time lapse} \\
\raisebox{2.5ex}{momenta:}&\raisebox{2.5ex}{$\bm\pi $} &\raisebox{2.5ex}{$\leftrightarrow$} &\raisebox{2.5ex}{$\displaystyle -\frac{\bom}{8\pi G}$} &
\raisebox{2.5ex}{:normal connection} \\
\raisebox{2.5ex}{rate of strain tensor:}&\raisebox{2.5ex}{$\bm\Sigma $} &\raisebox{2.5ex}{$\leftrightarrow$} &\raisebox{2.5ex}{$\displaystyle {\bm{\Theta_{\t}}}$} &
\raisebox{2.5ex}{:temporal extrinsic curvature} \\
\raisebox{2.5ex}{viscous stress tensor:}&\raisebox{2.5ex}{$\bm\Theta $} &\raisebox{2.5ex}{$\leftrightarrow$} &\raisebox{2.5ex}{$\displaystyle-\frac{\bm{\tilde\Theta_{\ts}}}{8\pi G}$} &
\raisebox{2.5ex}{:twisted radial extrinsic curvature} \\
\raisebox{2.5ex}{heat flux:}&\raisebox{2.5ex}{$\bm q $} &\raisebox{2.5ex}{$\leftrightarrow$} &\raisebox{2.5ex}{$\displaystyle-\frac{\rho^2\bom}{8\pi G}$} &
\raisebox{2.5ex}{:rescaled normal connection} \\
\raisebox{2ex}{external force:}&\raisebox{2ex}{$f_{A} $} &\raisebox{2ex}{$\leftrightarrow$} &\raisebox{2ex}{$-T_{\ts A}$} &
\raisebox{2ex}{:tangential matter stress } \\
\raisebox{2.5ex}{radiative heat transfer:}&\raisebox{2.5ex}{$\dot{Q}_{\mathrm{rad}} $} &\raisebox{2.5ex}{$\leftrightarrow$} &\raisebox{2.5ex}{$T_{\ts \t}$} &
\raisebox{2.5ex}{: matter radial flux } \\
\hline
\end{tabular}
\label{GF1}
\end{table}
\end{center}

We finally would like  to understand the radial constraint equation (\ref{Tss}) 
and compare it with the dynamical Young-Laplace equation \Ref{DLY}. 
Since the Young-Laplace equation describes a 2d interface embedded in 3 dimensions, 
it is convenient to introduce the spacelike metric 
\be
\bar{h}_{ab}= s_{a}s_{b} + q_{ab}.
\ee
This is the Riemannian metric of a 3d spacelike hyper-surface $\bar{\Sigma}$ on which $S$ is embedded.
The Gauss equation \Ref{Gauss} for this embedding reads 
\be
2 G_{\s\s}(\bar{h}) = \bm{\tilde{\Theta}}_{\s}:\bm{\Theta}_{\s}-R(q)~,
\ee
where $G_{\s\s}(\bar{h})= R_{\s\s}(\bar{h}) - \frac12 R(\bar{h})$ is the 3 dimensional Einstein tensor for $\bar h_{ab}$.
%not to be confused with 4d analog.
In the case of fluid dynamics we usually consider flat space metric, 
in which case this contribution vanishes. 
In our gravitational case it is easy to see that 
we can use this curvature contribution to redefine the matter radial pressure:
\be
\overline{T}_{\s\s} \equiv T_{\s\s} - \frac{G_{\s\s}(\bar{h})}{8\pi G}
\ee 
The radial constraint equation  can be simplified further by noticing that 
\be
(\rd +a)_{A}a^{A} = \frac{\Delta \rho}{\rho},
\ee
where $\Delta=\brd\dd\brd$ is the sphere Laplacian.
Therefore after rescaling by $\rho$, (\ref{Tss}) becomes 
\be
\boxed{
-(D_{\t} +\theta_{\t})\theta_{\n} = (8 \pi G) \left[  \overline{T}_{\ts\s} + \gamma \theta_{\s}\right]  
- \f12 \bm{\tilde{\Theta}}_{\t}:\bm\Theta_{\n} -\rho\bm{\om}^{2}
  - \Delta{\rho}}~,\label{Tsst}
  \ee
  where we have used the surface tension definition $\gamma =-\frac{\gamma_{\t}}{8\pi G}$.
 Let us first analyze this expression in the static case, which corresponds to an equilibrium configuration.
 This case corresponds to having no expansion $\theta_{\t}=0$, no rate of strain $ \bm{\Theta}_{\t}=0$ and no momenta flowing on the bubble so that $\bom=0$.
 In this case the equation simplifies to
 \be
 0=  \overline{T}_{\ts\s}+\gamma\theta_{\s}.
 \ee
 Since $\theta_{\s}$ is the radial curvature $\theta_S$ \Ref{b_theta},
 we recognize the Young-Laplace equation \Ref{LY}, 
 provided we identify $ \overline{T}_{\ts\s}$ with the pressure difference:
 \be
 P_{out}=  T_{\ts\s},\quad P_{in}= \f{G_{\ts\s}(\bar{h})}{8\pi G},\qquad \Delta P = \overline{T}_{\ts\s}.
 \ee
 the identification of $ T_{\ts\s}=\rho T_{\s\s}$ as the external pressure, which is the standard interpretation of the diagonal spatial component of the energy momentum  tensor.
 This dictionary is summarized in table 2.
 
\begin{center}
\begin{table}[ht]
\caption{Dictionary 2 for gravity-bubble thermodynamics correspondence}
\centering
\begin{tabular}{r c c c l}
\hline\hline
 Thermodynamical name  &   &  symbols &  & Gravity name \\ [0.5ex] % inserts table %heading
\hline\\
outside pressure :&$P_{out}$ &$\leftrightarrow$ &$T_{\ts\s}$ &
: matter pressure \\
\\
inside pressure :&$P_{in}$ &$\leftrightarrow$ &$\displaystyle
\frac{G_{\ts\s}(\bar{h})}{8\pi G} $ &
:radial 3d  Einstein tensor \\
\\
\raisebox{2.5ex}{ radial momenta discontinuity :}&\raisebox{2.5ex}{$\bar\pi_{r} $} &\raisebox{2.5ex}{$\leftrightarrow$} &\raisebox{2.5ex}{$\displaystyle \frac{\theta_{\n}}{8\pi G}$} &
\raisebox{2.5ex}{:timelike expansion}\\
\hline
\end{tabular}
\label{table:nonlin}
\end{table}
\end{center}
 When the equilibrium condition $ \gamma\theta_{\s}+ \Delta P=0$ is not satisfied, the 
 bubble is expanding by nucleation.
 If we assume that $\Theta_{\n AB}= \frac12 \theta_{\n} q_{AB}$ and that $ \bom=0$ which would be satisfied in a spherically symmetric situation, the equation for constant $\phi$ reduces to
 \be
 -\left(D_{\t} + \frac34 \theta_{\t}\right)\frac{\theta_{\n}}{8\pi G} = \Delta P + \gamma \theta_{\s}.
 \ee
%So we see that 
When the outside pressure dominates, $\Delta P + \gamma \theta_{\s}>0$, the bubble shrinks as it should in a fluid setting.
Thus, this equation suggests that $\theta_{\n}/(8\pi G)$ plays qualitatively the same role as the radial average momenta 
$\bar{\pi}_{r}=\int_{r_{-}}^{r_{+}} \rho v_{r}$ for a physical bubble\footnote{There is one equation on the literature that expresses the
rate of expansion of a bubble due to the difference between the outside pressure $P_{out}$ and the pressure $P_{\infty}$ at infinity, 
this is the Rayleigh-Plesset equation \ci{Plesset}. 
This equation is obtained by integrating radially the Navier-Stokes equation. 
In the case of a spherically symmetric bubble of radius $R$, expanding in an incompressible and non-viscous  fluid 
with constant density $\rho$, it is given by 
\be
\rho\left( \ddot{R}R + \f32 \dot{R}^{2} \right) = P_{out}- P_{\infty}
\ee
It shouldn't be confused with the equation  that expresses the rate of expansion 
by mass transfer due to the failure of the boundary equilibrium condition across the bubble surface $P_{out} + \gamma \theta_{\s} \neq P_{in}$.}.

 The full gravity equations \Ref{Tsst}  contains extra terms that can be written in the fluid language as terms proportional to $ \bm{\tilde{\Sigma}}:\bm\Sigma$ and $ \bm{q}\dd \bm{\pi} $.
 These represents dissipative terms controlling the bubble expansion  process.
 The relevance of these terms in the usual fluid setting is not established at this stage since there seems to be no known generalization of Young-Laplace equation for a viscous bubble.
 What's interesting is that the gravity analogy gives a proposal for what  such an equation
 should be.

%%%%%%%%%%%%%%%%%%%%%%%%%%%%%%%%%%%%%%%%%%%%%%%%%%%%%%%%%%%%%%%%%%%%%%%%%%%%%%%%%%%%%%%%%%%%%%
\section{Interfacial thermodynamics and black-hole thermodynamics}
 In the previous sections we have shown that the gravity equations projected onto  the gravitational screen 
 take the same forms as the non-equilibrium thermodynamic equations for a viscous bubble.
 In these three equations a common coefficient emerges, that is, the surface tension $\g$.
 It appears in the momenta conservation equation \Ref{momenta} as a negative 2-dimensional pressure \Ref{g=-p}, 
 in the generalized first law \Ref{u+phi} as a work term, 
 and in the radial equation \Ref{DLY} as the Young-Laplace extra pressure of curved boundaries.
 Comparing \Ref{Tsat}, \Ref{Tsnt} and \Ref{Tsst} with these confirms 
 that we can interpret the inward radial acceleration divided by $8\pi G$ as a surface tension, $\g=-\f{\g_{\t}}{8\pi G}$, 
 which has led to the consistent dictionary of tables 1 and 2. 
 This analogical interpretation seems at odd with the usual interpretation of 
 the radial acceleration that appears in the black-hole thermodynamics \ci{Hawking,Wald}. 
 In the black-hole analysis the outward radial acceleration times $\f{\hbar}{2\pi}$ is interpreted as a {\it temperature} and not as a surface tension. 
 Thus, there seems to be a tension between these two thermodynamic interpretations. 
 
 In this final section 
 we try to understand the black-hole thermodynamics from the surface-tension interpretation 
 and fill the gap. 
 The key idea here is to analyze the dependence of the surface tension on temperature. 
 We know that for most materials, when the temperature increases the surface tension diminishes. 
 This is governed by the {\it Gibbs equation}.
 Note however that the discussion in this section is much more speculative, 
 since we have not realized yet the physical identification of the surface tension 
 and a more general definition of temperature of the black hole, 
 which would require a discussion based on quantum mechanics. 

%%%%%%%%%%%%%%%%%%%%%%%%%%%%%
\subsection{Thermodynamics of interfacial systems: the Gibbs equation}
Let's review thermodynamics of an interfacial system \ci{Molecule, PCS}. 
We assume that we have an interface between two phases. 
For example, %a liquid-gas interface,
we consider a liquid-gas interface and denote the two phases by $L$ and $G$ 
while we represent quantities of the interface by index $I$. 
All the thermodynamic quantities such as the internal energy $U$ and total entropy $S$ 
decompose as a sum of ones associated with the liquid phase, the gas phase (the bubble) and the interface:
\be
U= U^{L}+ U^{G} + U^{I},\qquad S= S^{L}+ S^{G} + S^{I}.\nn
\ee
Suppose here that we are choosing Gibbs's dividing surface, in which the total volume is given by $V=V^{L} + V^{G}$, 
such that the total number of particle is described by 
\begin{equation}
N= N^{L } + N^{G}~,\nn
\end{equation}
since the fluid we are considering has only one component.
Then, the thermodynamic relation \Ref{interfU} becomes
\be\label{dUi}
\rd U^{I} = T\rd S^{I}  + \gamma \rd A~.
\ee
Here $U^I$ is a homogeneous, linear function of the extensive properties $S^I$ and $A$ of the system.
Therefore, we can integrate it with the intensive properties $T$ and $\g$ fixed (Euler's theorem) to obtain
\be\lb{U^i}
U^{I}= TS^{I} + \gamma A.
\ee
From here, we can define the interfacial free energy as $F^I \equiv U^I -TS^I = \g A~$ 
and its difference is $\rd F^I = -S^I \rd T + \g \rd A$. 
Then, we use the Maxwell relation for $F^I$ to get 
\begin{equation}\lb{Maxwell}
s^{I}\equiv \left( \f{\p S^I}{\p A}\right)_T=-\left( \f{\p \g}{\p T}\right)_A~,
\end{equation}
where $s^{I}$ represents the interfacial entropy area density.
In  a homogeneous system this evaluates to $s^I={S^I}/{A}$. 
Thus, we obtain the Gibbs equation:
\begin{equation}\lb{s^i}
s^I = -\left( \f{\p \g}{\p T}\right)_A~, 
\end{equation}
which means that in order to determine the surface entropy we need to measure how the surface tension changes with temperature. 
Furthermore, for an homogeneous system, the interfacial internal energy density is given by $u^I\equiv \f{U^I}{A}$,
so \Ref{U^i} and \Ref{s^i} lead to the thermodynamical equation of state: 
\begin{equation}\lb{u^i}
u^I = \g - T \left( \f{\p \g}{\p T}\right)_A~.
\end{equation}
This equation shows that the dependence $\g(T)$ can be understood as the equation of state for a one-component homogeneous interface. 
Especially, if the system has an internal energy density 
which is constant in $T$, $u^I_0=$const, 
we obtain by solving \Ref{u^i}
\begin{equation}\lb{g_const}
\g_0(T)=u^I_0 - s^I_0 T~, 
\end{equation}
which takes the same form as E\"otv\"os rule \ci{Eto}. Here $s^I_0$ is the constant entropy density.

%%%%%%%%%%%%%%%%%%%%%%%%%%%%%%%%%%%%%%%
\subsection{Black-hole thermodynamics from interfacial thermodynamics} 
We now try to interpret the black-hole thermodynamics from interfacial thermodynamics 
by putting some assumptions.  
We first remember that our gravity-thermodynamics correspondence has come from the fact that 
Israel's junction condition and the membrane paradigm enable us to describe 
dynamics of a bulk region by that of the surface energy-momentum tensor and surface metric on the screen. 
Then, the screen can be interpreted as a surface with energy and momentum embedded in the bulk spacetime, 
like a soap bubble in the air. 
Therefore, we can apply interfacial thermodynamics to the gravitational screen analogically, 
although the microscopic constituents are not identified yet. 
Note here that interfacial thermodynamics is based on the usual thermodynamical principle. 
Especially the additivity of extensive variables is used, 
in which long-distance interactions like gravitational force are not considered. 
Therefore, %in order to apply the general argument of interfacial thermodynamics to our gravity-thermodynamics correspondence, 
we \textit{assume} that we choose the time lapse $\rho$ 
for an observer $\t$ such that 
\begin{equation}\lb{rho_con}
\rho={\rm const.}
\end{equation}
Then, the gravitational terms in \Ref{Tsat}, \Ref{Tsnt}, and \Ref{Tsst} disappear, and 
the gravitational interaction between each part on the screen does not exist. 

Here a question arises:  
Where is the physical screen for a black hole? 
As discussed in the introduction, an event horizon is not a physical screen 
because it is teleological and null. 
Although we still don't obtain a complete definition of physical screens,  %which is local and timelike and satisifies the second law, 
we now consider, as a first trial, 
a timelike surface which just satisfies a condition that it is located near the event horizon 
of a stationary black hole. 
Because such a screen should have almost the same entropy of the stationary black hole, 
we \textit{assume} that the entropy density of the screen is given approximately by
\begin{equation}\lb{s_BH}
s^I_{BH}\approx \f{1}{4\hbar G}~. 
\end{equation}
%This is a constant value, and we can apply \Ref{g_const}. 
By using our dictionary in table \ref{GF1} and applying the Gibbs equation \Ref{g_const}, 
we obtain the temperature of the screen: 
\begin{equation}\lb{T_BH}
T_{BH}=\f{\hbar}{2\pi}(\g_{\t}-\theta_{\ts})~.
\end{equation}
Noting here that the screen is near the horizon, 
we \textit{assume} that the expansion $\theta_{\ts}$ is much smaller than $\g_{\t}$. 
Thus, the Hawking temperature \ci{Hawking} is approximately reproduced: 
\begin{equation}\lb{T_BH2}
T_{BH}\approx \f{\hbar}{2\pi}\g_{\t}~, 
\end{equation}
where $\g_{\t}$ is almost the same as the surface gravity of the stationary black hole. 

We discuss the meaning of this result. 
If we took the event horizon as the screen, $\theta_{\ts}$ would vanish and 
$T_{BH}= \f{\hbar}{2\pi}\g_{\t}$ would hold exactly. 
In the interfacial interpretation, however, it would have zero internal energy density $u^I=0$, 
which would not be physical. 
On the other hand, 
a timelike screen near the horizon has a small but finite internal energy density. 
Thus, combining the Bekenstein-Hawking entropy density \Ref{s_BH} 
and the Gibbs equation \Ref{g_const} reproduces 
the Hawking temperature \Ref{T_BH2}, which is consistent and non-trivial. 
Hence, there is a possibility that 
the temperature of a stationary back hole should be assigned more precisely by \Ref{T_BH} 
with a small correction term $-\f{\hbar}{2\pi}\theta_{\ts}$, 
although we need to understand it from a more fundamental point of view. 
%In this sense the interfacial understanding of the black-hole thermodynamics is expected to be more fundamental. 
%Note that since this discussion is based on the analogical and classical dictionary, 
%we need to investigate \Ref{s_BH} and \Ref{T_BH}
%from a microscopic and quantum-mechanical point of view, which will be done in future. 
%%%%%%%%%%%%%%%%%%%%%%%%%%%%%%%%
\subsubsection*{Examination of the assumptions} 
We here examine the assumptions above. %that have been put to derive \Ref{T_BH}. 
First, we note that the specific choice of the lapse \Ref{rho_con} can be realized 
at least locally, although it could not in the whole region of the screen.  
Next, the entropy density \Ref{s_BH} is the crucial and non-trivial input, 
but it could be justified by the entanglement entropy of spacetime \ci{Bianchi}. 
This condition also postulates through \Ref{g_const} %the dictionary and
that the internal energy density is constant on the screen. %, that is, the expansion $\theta_{\ts}$
It can be realized if we choose a screen located at a constant distance from the horizon 
and use the fact that the surface gravity of the Kerr-Newman black hole is constant. 
Furthermore, in order to reach \Ref{T_BH2} we assume that $\theta_{\ts} \ll \g_{\t}$. 
From the dictionary, 
this means that the 2d pressure is much larger than the internal energy density, 
which is not satisfied by the usual fluid. 
However, noting that the dominant energy condition can be broken by the Weyl anomaly \ci{KMY,KY}, 
such a situation could be possible in the near-horizon region. % with strong gravitational field. 
Thus, our discussion is speculative but suggestive. 
%Thus, our discussion is analogical and speculative 
%based on the assumptions that have been not established yet, 
%but it is suggestive 

%%%%%%%%%%%%%%%%%%%%%%%%%%%%%%%%%%%%%%%%%%%%%%%%%%%%%%%%%%%%%%%%%%%%%%%%%%%%%%%%%%%%%%%%%%
\subsection{Constituent equations}
We have written the gravity equations for a general screen in a thermodynamic manner. 
These equations are valid for any choice of screen's evolution because of diffeomorphism invariance. 
It is therefore natural to wonder whether  the choice of time evolution of the screen  
can also be encoded  thermodynamically. 
From a particular solution of Einstein's equation, 
we can get a solution of the thermodynamic equations for any choice of screen 
and its time evolution if the formulation is autonomous, which is not realized yet in this paper. 
%Given a gravity solution, 
It is here interesting to note that 
changing the time evolution of the screen amounts to change what are the  normal vectors to the screen.
This in turn changes the relation between $\bm\Theta_{\s}$ and $\bm\Theta_{\n}$. 
In our dictionary $\bm{\tilde{\Theta}}_{\s}$ is interpreted as the viscous stress tensor and $\bm\Theta_{\n}$ as the rate of strain tensor. 
Thus, we see that a change of the time evolution of a screen  corresponds to a {\it change of the constitutive laws}. 
Different screen's evolutions therefore correspond to different choices of materials or conditions under which a material is treated, 
which is still analogy just as discussed above. 

For instance, a lightlike evolution of a screen (e.g. event horizon) is obtained in the limit where $\n \to \s$. %although it is not a physical screen.  
Then, we have that  $\bm\Theta_{\n}=\bm\Theta_{\s}$, which corresponds to a Newtonian fluid \Ref{Newton_fluid}.
However, we have to be careful here. 
First, our dictionary in table \ref{GF1} and \ref{table:nonlin} would collapse because of $\rho=0$. 
Furthermore, the system would become unstable: 
The shear viscosity of this fluid is positive and equal to $1/(16\pi G)$, 
while the bulk viscosity is equal to $-1/(16\pi G)$, which is not physically reasonable. 
They agree with ones of the original membrane paradigm \ci{Price,Membrane,Carter}.  
Indeed, we have 
\be
\bm{\Theta} \equiv -\frac{\bm{\tilde{\Theta}}_{\ts}}{8\pi G} 
= -\frac{\bm{\tilde{\Theta}}_{\t}}{8\pi G} = -\frac{\theta_{\t}}{16\pi G}\bm q  + \frac{1}{16\pi G}2 \bm \sigma_{\t}~,\nn
\ee
where $\bm \sigma _{\t}$ denotes the traceless component of the rate of strain tensor. 
Hence, null evolution of screens cannot correspond to thermodynamic systems that satisfy the second law. % unless their expansion vanishes. 
%A typical example is an event horizon. %, which is not physical as we have discussed in the previous subsection. 
%This property is valid for an event horizon in a static space-time but not for an event horizon in a dynamical space-time, 
%since in this case the expansion does not vanish on the horizon, which is related to the teleological property of event horizon \ci{Membrane,Carter}.
In this way, from analogical point of view, we can consider 
time evolution of screens as that of continuum matters with the corresponding constituent equations.

%%%%%%%%%%%%%%%%%%%%%%%%%%%%%%%%%%%%%%%%%%%%%%%%%%%%%
\section{Conclusions and Discussions}
In this work we have seen how the Einstein gravity equations projected on a timelike membrane  
are equivalent to the thermodynamic equations for a viscous bubble.
These equations are of three kinds: 
the first one is the first law of thermodynamics that expresses the balance law of the internal energy, 
the second one is the Cauchy equation that expresses conservation of tangential momenta, 
which is also a viscous generalization of the Marangoni flow equation, 
and the third one is the dynamical generalization of Young-Laplace equation 
that expresses conservation of momenta across the bubble and governs its size due to evaporation or condensation.

In all these three equations one common parameter appears: the surface tension $\gamma$.  
%which is commonly given by the  the inward radial acceleration. 
It appears as a work term $\gamma \rd A$ in the first law, 
as a force term $\brd \gamma$ in the Marangoni equation 
and as a pressure term $\gamma \theta_{S}$ in the Young-Laplace equation.
Finally the Gibbs equation determines  the interfacial entropy density $s^I$ 
from the temperature dependence of the surface tension as $\left(\f{\partial\gamma}{\partial T}\right)_{A} = -s^I$.

%The identification of a  gravitational screen  as a fluid interface allows us to identify 
%a full dictionary between gravity and thermodynamics.
The non-trivial and consistent interpretation of the surface tension among three equations 
allows us to identify a consistent analogical dictionary between gravity and thermodynamics. 
In the dictionary the surface tension is proportional to the inward radial acceleration 
while the internal energy is proportional to the inward radial expansion. 
We can also assign to the gravitational screen a viscous stress tensor, 
a rate of strain tensor, momenta, a Newtonian potential  and a heat flux vector (see tables \ref{GF1} and \ref{table:nonlin}).
Furthermore, using the Gibbs equation and the correspondence between the choice of the screen time evolution and that of constituent equations, 
we can interpret the usual black-hole thermodynamics in terms of this dictionary 
as an analogy. 

In order to go beyond this analogy  
we need to get a deeper understanding of the equation of states and of the constituent equations.
In order to devise an  equation of states one would need to find, for an arbitrary screen, a notion of temperature. 
This should appear in a quantum treatment of the gravitational field as suggested by Unruh's work \ci{Unruh} 
and along the lines of \ci{matteo}. 
Here we could make a speculative discussion for a more general definition of temperature. 
Consider a screen with constant internal energy density. %in which the entropy density is also constant. 
If we postulate again that $s^I=\f{1}{4\hbar G}$, %which could be motivated by the entanglement entropy of spacetime \ci{Bianchi}, 
then, we would obtain the temperature which takes the same form as \Ref{T_BH}. 
Unlike the case of a screen near the horizon, generally, 
$\theta_{\ts}$ is not small and could bring a very different value of the temperature from Unruh's one. 
Such an attempt %, which is motivated by interfacial thermodynamics of screens, 
could introduce a more general idea of temperature which takes dynamics of spacetime itself into consideration. 

We have identified the entropy production term of the gravitational screen, 
which is proportional to $-\bm{\Theta}_{\n}\!: \! \bm {\tilde{\Theta}}_{\s}$.
One of the central challenges that lie ahead is to understand, describe and characterize 
for which screens this quantity is always positive.
Such screens would satisfy the second law, and then, 
the analogy that we have been developing here would become a physical correspondence.
This is the first central question to be resolved next.
Understanding the second law also necessitates to understand geometrically 
the relationship between the heat flux vector to the temperature gradient. 

If it can be shown that it is always possible to choose the time evolution of the screen 
in agreement with the second law, 
we can then start to answer our initial question about the entropy of gravitational system. 
Such  possibility would naturally suggest a consistent picture for gravitational screen 
in agreement with the old dream of holography.

%%%%%%%%%%%%%%%%%%%%%%%%%%%%%%%%%%%%%%%%%%%%%%%%%%%%%%%%%%%%%%%%%%%%%%%%%%%%%%%%%%%%%%%%%%%%
\section*{Acknowledgements}
We are  grateful to E. Bianchi, D. Minic, L. Lehner, T. Jacobson, M. Smerlak and H. Haggard  for many insight-full exchanges.
And we would like to thank T. Araki and A. Onuki for helpful comments about surface tension.
%We thank T. Araki and A. Onuki for helpful comments about surface tension.
This work has been refined by the audience of colloquiums L.F. gave in Virginia Tech, 
Illinois University and Perimeter institute, and we thank the audience for their questions.

This work started while L.F. was a visitor at Yukawa institute in Kyoto, he thanks deeply N. Sasakura for an inspiring stay in this magnificent city.
This research was supported in part by Perimeter Institute for Theoretical Physics. Research at Perimeter Institute is supported by the Government of Canada through Industry Canada and by the Province of Ontario through the Ministry of Research and Innovation. This research was also partly supported by grants from NSERC. 

Y. Y. was supported by Grant-in-Aid of the MEXT Japan for Scientific Research (No. 25287046) 
and for the Global COE Program ``The Next Generation of Physics, Spun from Universality and Emergence'', 
and by the JSPS Research Fellowship for Young Scientists.  
%and by the Grant-in-Aid for the Global COE Program from the MEXT of Japan.% ``The Next Generation of Physics, Spun from Universality and Emergence'' 
%from the Ministry of Education, Culture, Sports, Science and Technology (MEXT) of Japan.
He thanks Perimeter Institute for the Visiting Fellowship program.% and hospitality. 
%%%%%%%%%%%%%%%%%%%%%%%%%%%%%%%%%%%%%%%%%%%%%%%%%%%%%%%%%%%%%%%%%%%%%%%%%%%%%%%%%%%%%%%%%
\appendix
\section{Gauss Codazzi and Ricci equation}
%In this section 
We present the equations that govern the embedding of a hyper surface in space-time \ci{Poisson}.
We denote by $\bm g$ the space-time metric and by $\n$ the normal to the hyper surface, 
whose signature is given by $\epsilon = \n\dd\n$. 
The induced metric $\bm h$ on the hyper surface and its  extrinsic tensor $\bm K$ are given by
\be
h_{ab}\equiv g_{ab}-\epsilon n_{a}n_{b},\qquad K_{ab} \equiv h_{a}{}^{\bar{a}}h_{b}{}^{\bar{b}}\nabla_{\bar{a}}n_{\bar{b}}.
\ee 
We also introduce 
\be
\widetilde{K}_{ab}\equiv K h_{ab} - K_{ab},
\ee
where $K=h^{ab}K_{ab}$.
Our conventions are such that if $\nabla_{a}$ denotes the covariant derivative preserving $\bm g$ 
its Riemann tensor is defined as 
\be
[\nabla_{a},\nabla_{b}] v^{c}=R_{ab}{}^{c}{}_{d}(g) v^{d}.
\ee

The contracted Codazzi equation is given by
\be\label{Codazzi}
h_{a}{}^{b}G_{b \n}(g) = - D_{b}\widetilde{K}^{b}{}_{a}~,
\ee
where $G_{ab}(g)= R_{ab}(g) - \f12 g_{ab} R(g)$ is the Einstein tensor for $\bm g$, and 
$D_a$ is the surface covariant derivative defined by $D_a V^b\equiv h_a{}^{\bar a}h^b{}_{\bar b} \N_{\bar a} V^{\bar b}$ for $h^a{}_bV^b = V^a$.
The contracted Gauss equation is given by
\be\label{Gauss}
2 G_{\n\n}(g) =  \bm{\widetilde{K}}: \bm{K}  -\epsilon R(h)~,
\ee
where $R(h)$ is the Ricci scalar for $\bm h$. 
We finally consider the Ricci identity that relates the two curvature tensors
\be\label{Ricci}
R(g) = R(h) + \eps \bm{\widetilde{K}}: \bm{K} - 2 \nabla_{a}a_{\n}^{a} - 2\eps \nabla_{a}(n^{a}K)~,
\ee
where $a_{\n}^{a} \equiv -\epsilon \nabla_{\n}n^{a}$ is the acceleration.

%%%%%%%%%%%%%%%%%%%%%%%%%%%%%%%%%%%%%%%%%%%%%%%%%%%%%%%%5
\section{Boundary evolution}
Here we derive the dynamical Young-Laplace equation \Ref{DLY}. 
We suppose the interface as the thin limit of a finite layer between two phases. 
That is, if the size of the interface is $\Delta r=r_+-r_-$, the thin limit means $\Delta r \rightarrow 0$. 
Here $r$ is the distance from the interface located at $r=0$ after taking the limit, 
and $r_+,~r_-$ are the positions of the outside and inside layer, respectively.
Note that generally the surface is moving with the velocity $\dot r$. 
In this choice of coordinates $s^i \p_i =\p_r$ holds where $\s$ is the normal vector 
to the interface. Therefore for this coordinates we can assume that $V_{\s}= V_{r}$.
We also denote by $\theta_S=\pa_{i}s^{i}$ the curvature of the interface.
 Given a physical quantity $f$ we define its average across the surface to be
 \be
\bar{f} \equiv \int_{r_{-}}^{r_{+}} f  \rd r,
\ee
 where $\Delta r$ is taken to be infinitesimal.
We now construct a useful formula which expresses the conservation across the interface.
For a vector $V^i$ its divergence can be evaluated by using $\delta_{ij}=q_{ij}+s_is_j$ as 
\bea
\p_i V^i  &=& q_i{}^k \p_k (q^i{}_j V^j)+s_i \p_r(q^i{}_j V^j)+q_i{}^k\p_k(s^iV_{\s})+s^i\p_r(s^iV_{\s})\nn\\
 &=&\rd_AV^A+s_i\p_r(q^i{}_jV^j)+\theta_S V_{r} +\p_r V_{r}~.\nn
\eea
Integrating this and taking the thin limit we get
\bea\lb{formula}
\int_{r_-}^{r_+}\rd r \p_i V^i 
&\longrightarrow & \rd_A \int_{r_-}^{r_+}\rd r V^A + s_i \int_{r_-}^{r_+}\rd r \p_r(q^i{}_j V^j) + \theta_S \int_{r_-}^{r_+}\rd r V_{\s} + \int_{r_-}^{r_+}\rd r \p_r V_{\s} \nn\\
 &=&\rd_A \overline{V}^A + \theta_S \overline{V}_{r} + [V_{r}]^+_-~,
\eea
where the orthogonality $s_i q^i{}_j|_{\pm }=0$ and the definition of the curvature $\theta_S \equiv q^{ij}\p_i s_j=\p_i s^i$ have been used.
The bracket $[V]^{+}_{-}= V(r_{+}) - V(r_{-})$ is the discontinuity.
If one applies this to the mass conservation equation 
$
 \pa_{t}\rho + \pa_{a}(\rho v^{a}) =0,
 $ and use the continuity of the tangential 
velocity vector $v^{A}$ across the interface 
we obtain 
\be
\pa_{t}\bar{\rho} + \rd_{A}(\bar\rho v^{A}) = [\rho (\dot{r}- v_{r})]_{-}^{+}
 -  \theta_S \bar{\pi}_{r}.
 \ee
 We see that $ j_{+} \equiv \rho^{+} (v_{r}^{+}- \dot{r})$ represents the mass flux per unit area  towards the region $r>r_{+}$
 and $\dot{m}_{+}=\int_{S} j_{+} \rd A$ represent the change of mass of this region.
 Similarly $j_{-}\equiv -\rho^{-}(v_{r}^{-}- \dot{r})$ represents the mass flux outside the region $r<r_{-}$ and $ \dot{m}_{-}=- \int_{S} j_{-} \rd A$
 represent the change of mass of this region.
 We see also  that if the normal momenta surface density does not vanish we also have an additional  term proportional to the curvature.
 In the following we demand the continuity of the velocity field $[v_{i}]_{-}^{+}=0$.
 Under this condition the  conservation of mass of the interface reads 
 \be\label{masscons}
 [\rho]_{-}^{+} (\dot{r}- v_{r})
 -  \theta_S \bar{\pi}_{r}=0.
 \ee

Now we derive time evolution equation of the bubble radial momentum density 
$
\bar \pi_r \equiv \int_{r_-}^{r_+}\rd r \rho v_r,
$
from the $r-$component of the momentum conservation law: 
\begin{equation}
%\rd_t (\rho v_r) + (\p_i v^i) \rho v_r = \p_j T^j{}_r 
%\Longleftrightarrow 
\p_t(\rho v_r) + \p_i (\rho v_r v^i) = \p_j T^j{}_{\s}~,
\end{equation}
where $T_{ij}$ is the total stress tensor.
Using this equation of motion and the thin-limit formula \Ref{formula}, we obtain that 
\be
\p_t \bar \pi_r + \rd_A (\bar \pi_r v^A) = [\rho v_r (\dot r-v_{r})]^+_-
-\theta_S \overline{\rho v_{r}^{2}} + \rd_{A} \overline{T}^{A}{}_{\s}
+ [T_{rr}]_{-}^{+} + \theta_S \overline{T}_{rr}.
\ee 
The first two terms in the RHS cancel each others due to the
 mass conservation equation (\ref{masscons}) and the continuity of $v_{r}$.
 ${[}T_{rr}]_{-}^{+}$ gives the difference of dynamical radial pressure: $T_{rr}=-P$.
 Finally the expectation value of the stress tensor contains the definition of the surface tension
 \be\label{surf}
 \overline{T}^{ij} = \gamma s^{i}s^{j}.
 \ee
 Given these we see that the previous thin limit equation reduces to the dynamical Young-Laplace equation
 \begin{equation}
\p_t \bar \pi _r + \rd_A (\bar \pi_r v^A) = -(\Delta P + \g \theta_S) ~,
\end{equation}
 This is the dynamical Young-Laplace equation \Ref{DLY}.

 Concerning the point (\ref{surf})
note here that the stress tensor contains the effect of the surface tension \ci{Onuki}
via 
\begin{equation}
T_{ij}^{surface}=-M\p_i n \p_j n~,
\end{equation}
where $n$ is the number density, and $M(n)$ is a function of $n$, which comes from the gradient term 
in the Ginzburg-Landau effective free energy \ci{Landau_S}. 
Only this term contributes to the thin limit, and its integration gives the surface tension: 
\begin{equation}
\int_{r_-}^{r_+}\rd r (-T_{ij}^{surface})s^i s^j =  \int_{r_-}^{r_+}\rd r M (\p_r n)^2 = \g~,
\end{equation}
which is the same as the original theory by van der Waals \ci{Waals}.

%%%%%%%%%%%%%%%%%%%%%%%%%%%%%%%%%%%%%%%%%%%%%%%%%%%%%%%%%%%%%%%%%%%%%%%%%%%%%%%%%%%%%%%%%%%%%%%%%%%%%%%%%%%%%%%%%%%%%%%%%%%%%%%%%%%%%%%%%%%%%%%%%%%%%%%%%
%\section*{References}

\end{document}